\documentclass[sigconf, screen]{acmart}

\usepackage{tabu}
\usepackage{listings}
\usepackage{bm}
\usepackage{float}
\usepackage[caption=false]{subfig}
\usepackage{amsmath}
\usepackage{amssymb}
\usepackage{multirow}
\usepackage{balance}

\setcopyright{acmlicensed}
\acmPrice{15.00}
\acmDOI{10.1145/3213846.3213858}
\acmYear{2018}
\copyrightyear{2018}
\acmISBN{978-1-4503-5699-2/18/07}
\acmConference[ISSTA'18]{27th ACM SIGSOFT International Symposium  on  Software Testing and Analysis}{July 16--21, 2018}{Amsterdam, Netherlands}

\begin{document}
\title[Identifying Implementation Bugs in Machine Learning based Image Classifiers ...]{Identifying Implementation Bugs in Machine Learning Based Image Classifiers using Metamorphic Testing}

\author{Anurag Dwarakanath}
\affiliation{%
  \institution{Accenture Technology Labs}
  \city{Bangalore} 
  \country{India} 
}
\email{anurag.dwarakanath@accenture.com}

\author{Manish Ahuja}
\affiliation{%
  \institution{Accenture Technology Labs}
  \city{Bangalore} 
  \country{India} 
}
\email{manish.a.ahuja@accenture.com}

\author{Samarth Sikand}
\affiliation{%
  \institution{Accenture Technology Labs}
  \city{Bangalore} 
  \country{India}}
\email{s.sikand@accenture.com}

\author{Raghotham M. Rao}
\affiliation{%
  \institution{Accenture Technology Labs}
  \city{Bangalore}
  \country{India}
}
\email{raghotham.m.rao@accenture.com}

\author{R. P. Jagadeesh Chandra Bose}
\affiliation{%
  \institution{Accenture Technology Labs}
  \city{Bangalore}
  \country{India}
}
\email{jagadeesh.c.bose@accenture.com}

\author{Neville Dubash}
\affiliation{%
  \institution{Accenture Technology Labs}
  \city{Bangalore}
  \country{India}
}
\email{neville.dubash@accenture.com}

\author{Sanjay Podder}
\affiliation{%
  \institution{Accenture Technology Labs}
  \city{Bangalore}
  \country{India}
}
\email{sanjay.podder@accenture.com}

\renewcommand{\shortauthors}{A. Dwarakanath et al.}

\begin{abstract}
We have recently witnessed tremendous success of Machine Learning (ML) in practical applications. Computer vision, speech recognition and language translation have all seen a near human level performance. We expect, in the near future, most business applications will have some form of ML. However, testing such applications is extremely challenging and would be very expensive if we follow today's methodologies. In this work, we present an articulation of the challenges in testing ML based applications. We then present our solution approach, based on the concept of Metamorphic Testing, which aims to identify implementation bugs in ML based image classifiers. We have developed metamorphic relations for an application based on Support Vector Machine and a Deep Learning based application. Empirical validation showed that our approach was able to catch 71\% of the implementation bugs in the ML applications.  
\end{abstract}

%
%



\maketitle

\section{Introduction}
Software verification is the task of testing whether the program under test (PUT) adheres to its specification \cite{IEEEStdVV}. Traditional verification techniques build a set of `test cases' which are tuples of (input - output) pairs. Here, a tester supplies the `input' to the PUT and checks that the output from the PUT matches with what is provided in the test case. Such `expected output' based techniques for software testing are widespread and almost all of the white-box and black-box techniques use this mechanism.

\begin{figure} [b] 
	\includegraphics[height = 1.9in]{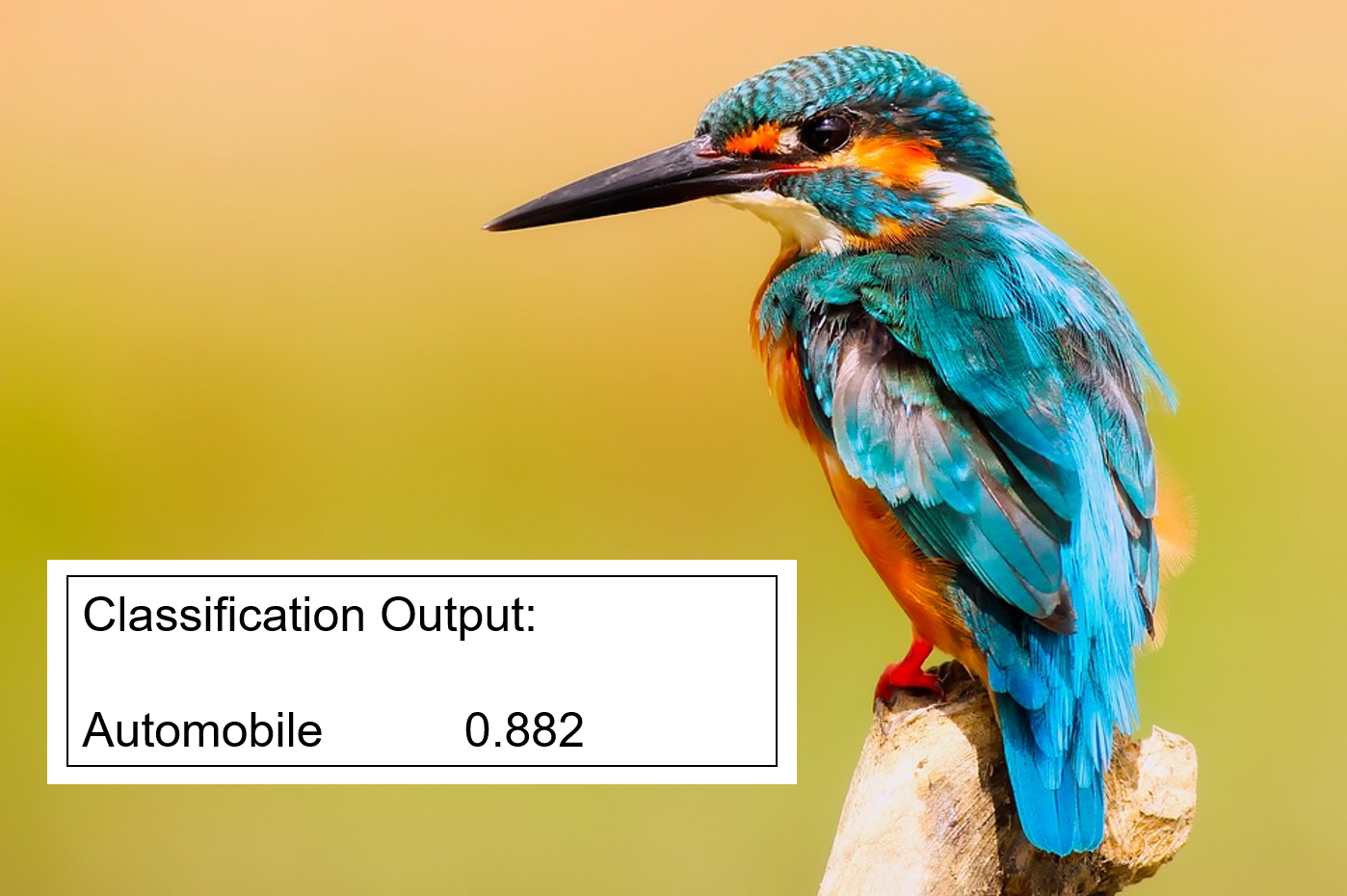}
\caption{An example input to \& output from a ML based image classifier. The classification is clearly wrong from a human-oracle expectation. Problem definition: Software verification should be able to ascertain that a) the classification is correct as per the specification (i.e. for the given training data, the classification for the example is correct); b) score of 0.882 for the class is correct. Image source: \cite{birdImage}}
\label{probDef}
\end{figure}

However, when a machine learning (ML) based application comes to an independent testing team (as is typically the case before `go-live'), verifying the ML application through (input-output) pairs is largely in-feasible. This is because:

\begin{enumerate}
	\item The PUT is expected to take a large number of inputs. For example, an image classifier can take \textbf{any} image as its input. Coming up with all the scenarios to cover this large possibility in inputs would be too time consuming;
	\item In many cases, the expected output for an input is not known or is too expensive to create. See Figure \ref{probDef} for an example.
	\item Unlike testing of traditional applications, finding one (or a few) instances of incorrect classification from an ML application does not indicate the presence of a bug. For example, even if an image classifier gives an obvious wrong classification, a `bug report' cannot be created since the ML application is not expected to be 100\% accurate. 
\end{enumerate}

The current options for a tester to test ML applications is largely left to validation. The tester would acquire a large number of real-life inputs and check that the outputs meet the expectation (with the tester acting as the human oracle). Such validation based testing would be very expensive in terms of time and cost. 

When a ML algorithm exhibits an incorrect output (e.g. wrong classification to a particular input), there can be multiple underlying reasons. These include - a) deficient (e.g. biased) training data; b) poor ML architecture used; c) the ML algorithm learnt a wrong function; or d) there is an implementation bug. In the current state of practice, almost always an incorrect output is attributed to deficient training data and the developers would be urged to collect more diverse training data. However, we believe, the first aspect to ascertain is that the ML algorithm does not have implementation bugs. For example, if the fault that was seen is due to an implementation issue, getting more training data will not help.

`Metamorphic Testing' \cite{chen1998metamorphic} is a prominent approach to test an ML application. Here, a test case has two tuples of ($\textrm{input}_1-\textrm{output}_1$) and ($\textrm{input}_2-\textrm{output}_2$). The second input, $\textrm{input}_2$, is created such that we can reason about the relation between $\textrm{output}_1$ and $\textrm{output}_2$. This relation between the outputs is termed as a Metamorphic Relation (MR). When we spot cases where the relation is not maintained, we can conclude that there is a bug in the implementation.

In this paper, we show how a tester can efficiently (with one or a few test cases) identify implementation issues in ML applications. Our approach is based on the concept of Metamorphic Testing and we have worked with two publicly available ML applications for image classification. The first application uses a Support Vector Machine (SVM) with options of a linear or a non-linear kernel. The second application is a deep learning based image classifier using a recent advancement in Convolutional Neural Networks, called a Residual Network (ResNet). For both the applications, we have developed metamorphic relations (MRs) and we provide justifications with proofs (where applicable), examples and empirical validation. 

Our solution approach was validated through the concept of Mutation Testing \cite{jia2011analysis}. Bugs were artificially introduced into the two ML applications and we checked how many of these bugs could be caught through our technique. The tests showed that 71\%  of the implementation bugs were caught through our approach.

Our work advances the research in Metamorphic Testing in multiple ways: (i) we investigate the case of an SVM with a non-linear kernel, which hasn't been done before, and we develop a new metamorphic relation (MR) for the linear-kernel; (ii) our work is the first to provide formal proofs of the MRs based on the mathematical formulation of SVM; (iii) we also develop MRs for the verification of a deep learning based image classifier, which is the first such work; and (iv) we evaluate the goodness of our approach through rigorous experiments and all of our data, results \& code are open-sourced \footnote{\url{https://github.com/verml/VerifyML}}. 



We anticipate our approach to be used by a tester as follows: when a ML based application is submitted to his/her desk for testing, the first step of verification can be done in an automated fashion through our approach. The tester would select relevant metamorphic relations and a tool based on our approach would automatically create test cases to check for the properties. If any of the properties do not hold, the verification has identified an implementation bug in the ML application.

This paper is structured as follows. We present the related work in Section \ref{relatedWork}. Section \ref{ourApproach} presents our approach for two ML based image classifiers. Section \ref{results} details the experimental results and we conclude in Section \ref{conclusion}.

\section{Related Work} \label{relatedWork}


Typical testing activity includes building tuples of (input-expected output). However, in many cases, the expected output cannot be ascertained easily \cite{xie2011testing}. Examples include scientific programs which implement complex algorithms, programs which try to predict answers where the correct answer is not known (e.g., a program trying to predict the weather \cite{murphy2008improving}). Such programs have been termed as `non-testable programs' \cite{weyuker1982testing} and machine learning algorithms are considered to be in this class.

Various methods have been devised to test `non-testable programs'. These include constructing an oracle from formal specification \cite{barr2015oracle} (and recently formal verification has been attempted for machine learning \cite{selsam2017developing}), developing a pseudo oracle (or multiple implementations of a program \cite{srisakaokul2018multiple}), testing on benchmark data \cite{resNetDefaultChoice}, assertion checking \cite{murphy2009using}, using earlier versions of the program, developing trivial test cases where the oracle is easily obtainable \cite{murphy2010empirical}, using implicit oracles (such as when a program crashes) and Metamorphic Testing \cite{chen1998metamorphic}. While none of these methods have been found to solve all aspects \cite{barr2015oracle}, Metamorphic Testing has been found to be quite useful \cite{liu2014effectively}. We refer the readers to a recent survey \cite{barr2015oracle} for a discussion on these various methods.

There has been work in the application of Metamorphic Testing over machine learning algorithms. This includes the testing of Naive-Bayes classifier \cite{xie2009application}\cite{xie2011testing}, support vector machine with a linear kernel \cite{murphy2010empirical} and k-nearest neighbor \cite{xie2009application} \cite{xie2011testing}. However, none of these works investigate the case of an SVM with non-linear kernel (which is commonly used). Our work also applies metamorphic testing on deep learning based classifiers. 


A recent work on applying Metamorphic Testing to test deep learning frameworks has been made \cite{ding2017validating}. However, this work attempts to perform `validation' and develops MRs based on the typical expectations from deep learning. For example, the first MR developed in \cite{ding2017validating} claims that the classification accuracy of a deep learning architecture should be better than a SVM. Further, the work in \cite{ding2017validating} does not provide justifications for the relations and does not perform an empirical validation.

Another recent work in the verification of ML is the use of formal methods \cite{selsam2017developing}. This work has developed a  mathematical language and an associated tool (called Certigrad) where a developer can specify the ML algorithm and verify the implementation using an interactive proof assistant. However, a key challenge of this method is the scalability to support a wide range of machine learning methods, the need for developers to learn a new formal language and the inability to check whether the specification written in the formal language is correct. Nevertheless, this work recognizes the difficulty of verifying machine learning algorithms and focuses on the case of identifying implementation bugs (which our work does as well). 

Finally, some recent work has been made to specifically identify failure cases in deep learning architectures \cite{tian2017deeptest} \cite{pei2017deepxplore}. These works aim to `validate' a deep learning system by building test inputs reflecting real-world cases. The focus of such work is in the identification of deficiencies in training data, or using a poor learning algorithm. Similar work is done by the entire area of adversarial examples \cite{papernot2016practical}. In contrast, our work specifically investigates the case of `verification' where the underlying root cause for a fault is an implementation bug.

\section{Identifying Implementation Bugs in ML based applications} \label{ourApproach}
In this section, we will present our solution methodology. We have designed the metamorphic properties and validated our approach on the following publicly available ML based applications.

\begin{itemize}
    \item Hand-written digit recognition from images using SVM with a linear and a non-linear kernel \cite{SVMAppln}
    \item Image classification using Residual Networks (a type of Deep Convolutional Neural Network) \cite{resNetImplementation}
\end{itemize}

\subsection{Metamorphic Relations for Application 1: Digit Recognition using SVM }

We selected an application \cite{SVMAppln} that classifies images of hand written digits into classes of $0$ to $9$. The application is an official implementation from Scikit-learn, a popular machine learning library and uses a Support Vector Machine (SVM) for the classification. The application can be configured to use either a linear kernel or a non-linear kernel. We have developed metamorphic relations for both configurations with the non-linear kernel being an RBF kernel. 

%

For training, the SVM application takes labeled examples of hand-written digits and learns decision boundaries separating the classes. Each example is an image of 8 pixels by 8 pixels. The pixels are in gray-scale and one example is represented by an array of 64 numbers. See Figure \ref{permFeatures}(a) for visualization of some of the examples. 

For testing, the application takes an example and predicts the class (between `0' \& `9'). The application also provides a score depicting its confidence on the classification. The score is the (functional) distance of the example from the decision boundary - a large score means the example is farther away from the decision boundary and the classifier is more certain of its decision. 

We have developed the following metamorphic relations (MRs): 

\begin{enumerate}
    \item MR-1: Permutation of training \& test features
    \item MR-2: Permutation of order of training instances
    \item MR-3: Shifting of training \& test features by a constant (only for RBF kernel)
    \item MR-4: Linear scaling of the test features (only for linear kernel)
\end{enumerate}

To justify the validity of the MRs, we will provide proofs based on the formulation of the SVM algorithm.

The SVM application uses the formulation of the LIBSVM \cite{libsvm} library. The formulation can be represented in primal form and in a corresponding dual form. The optimization is done, by default, in the dual form and our proofs are also based on this form. 

Let the training dataset be represented by $(x_i, y_i)$, $i = 1, . . . , m$, $x_i \in R^n$, $y_i \in  \{-1, +1\}$. $x_i$ denotes one hand-written image and $y_i$ is its label. LIBSVM solves the following constrained optimization problem to find the optimal value of $\alpha$ (this is the dual form).
\begin{equation}
	\begin{aligned}
		& \underset{\alpha}{\text{minimize}}
		& & \frac{1}{2}{\alpha}^TQ\alpha - {e}^T\alpha \\
		& \text{subject to}
		& & y^T\alpha = 0, 0 \leq {\alpha}_i \leq C , \forall i = 1 \ldots m  
	\end{aligned}
	\label{svmDual}
\end{equation}
Here, Q is a square matrix whose each element (i, j) is defined as: $Q_{ij} = y_{i}y_{j}K(x_{i}, x_{j})$. The function $K(x_{i}, x_{j})$ is the kernel. Once the optimization is complete, given a test instance $x_a$, the functional distance of the instance from the decision boundary is given by Equation (\ref{funcDist}). The class is $1$ if the functional distance is $\geq 0$ else $-1$.

\begin{equation}
	\begin{aligned}
		D(x_a) = \sum_{i=1}^{i=m} \alpha_i y_i K(x_a, x_i) + b
	\end{aligned}
\label{funcDist}
\end{equation}
For the linear kernel, we have $K(x_{i}, x_{j}) = x_{i}^{T}x_{j}$ (i.e. the dot-product between the data instances $x_{i}$ and $x_{j}$). For the RBF kernel we have $K(x_{i}, x_{j}) = e^{-\gamma ||x_{i} - x_{j}||^2}$ (i.e. a measure of the distance between the two data instances $x_{i}$ and $x_{j}$). 

Note that the formulation of SVM in dual form is completely expressed by the kernel function $K(x_i, x_j)$. This forms the underlying principle of our proofs where we show that transforming $x_i$ and $x_j$ in a particular way will lead to the transformation of the kernel $K(x_i, x_j)$ in a known and decidable way.

\subsubsection{MR-1: Permutation of Training \& Test Features} \label{svmMR1}
Let $X_{train}$ be the set of training data and $X_{test}$ be the set of test data. Upon completion of training via SVM, let a particular test instance, $x_{test}^i$ be classified as class $c$ with a score of $s$. MR-1 specifies that if we re-order the features of $X_{train}$ and $X_{test}$ through a deterministic function, say $perm()$, and re-train the SVM with $perm(X_{train})$, then the test instance $perm(x_{test}^i)$ will continue to be classified as class $c$ with the same score of $s$.

\begin{figure}
	\subfloat[Original training data]{%
	\includegraphics[height=1.5cm]{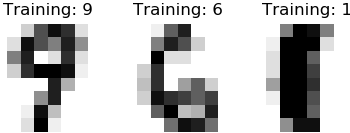}}%
	\hspace{18pt}%
	\subfloat[Original test data]{%
	\includegraphics[height=1.5cm]{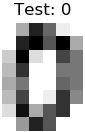}}%
	\\
	\subfloat[Permuted training data]{%
	\includegraphics[height=1.5cm]{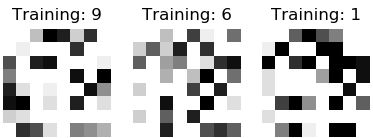}}%
	\hspace{18pt}%
	\subfloat[Permuted test data]{%
	\includegraphics[height=1.5cm]{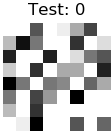}}%
	\caption{Visualization of MR-1. The results would be the same whether the SVM is trained \& tested on the original or permuted data.}
	\label{permFeatures}
\end{figure}

To give an example of the function $perm()$, consider two data instances $x_a = (x_1^a, x_2^a, x_3^a)$ and $x_b = (x_1^b, x_2^b, x_3^b)$. The function $perm()$ re-orders the features as follows: $x_{a}^{p} = (x_2^a, x_3^a, x_1^a)$ and $x_{b}^{p} = (x_2^b, x_3^b, x_1^b)$ (note that both the instances are permuted in the same way). The feature permutation is visualized in Figure \ref{permFeatures}.

\begin{proof}
Consider the examples $x_a$ and $x_b$ and their permuted versions $x_{a}^{p}$ and $x_{b}^{p}$. From Equation (\ref{svmDual}), the impact of the permutation is in the kernel function $K(x_a, x_b)$. For the linear kernel we have $K(x_a, x_b) = {x_a}^T {x_b} = x_1^a x_1^b + x_2^a x_2^b + x_3^a x_3^b$. Further, $K(x_a^{p}, x_b^{p})= {x_a^{p}}^T {x_b^{p}} = x_2^a x_2^b + x_3^a x_3^b + x_1^a x_1^b = K(x_a, x_b)$. 

Similarly, for the RBF kernel we have:

$K(x_a, x_b) = e^{-\gamma ((x_1^{a} - x_1^{b})^2 + (x_2^{a} - x_2^{b})^2 + (x_3^{a} - x_3^{b})^2)} $. Further, $K(x_a^{p}, x_b^{p})= e^{-\gamma ((x_2^{a} - x_2^{b})^2 + (x_3^{a} - x_3^{b})^2 + (x_1^{a} - x_1^{b})^2)} = K(x_a, x_b)$. 
\end{proof}

\subsubsection{MR-2: Permutation of Order of Training Instances}
Let $X_{train}$ be the set of training data. This MR specifies that if we shuffle $X_{train}$ (i.e. change the order of the individual examples), the results would not change. This is evident from Equation (\ref{svmDual}) where re-ordering the training instances, re-orders the constraints. However, the optimization is done to satisfy all the constraints and therefore the constraints' order will make no difference.

\subsubsection{MR-3: Shifting of Training \& Test Features by a Constant (only for RBF Kernel)}
Let $X_{train}$ be the training data \& $X_{test}$ be the test data. Upon training via SVM, let a particular test instance, $x_{test}^i$ be classified as class $c$ with a score of $s$. MR-3 specifies that if we add a constant $k$ to each feature of $X_{train}$ \& $X_{test}$, the re-trained SVM will continue to classify $x_{test}^i$ as class $c$ with score $s$.

\begin{proof}
Consider the examples $x_a = (x_1^a, x_2^a, x_3^a)$, $x_b = (x_1^b, x_2^b, x_3^b)$ and their shifted versions $x_a^s = (x_1^a + k, x_2^a + k, x_3^a + k)$ and $x_b^s = (x_1^b + k, x_2^b + k, x_3^b + k)$. For the RBF kernel we have, 

$K(x_a^s, x_b^s) = e^{-\gamma ((x_1^{a} + k - (x_1^{b} + k))^2 + (x_2^{a} + k - (x_2^{b} + k))^2 + (x_3^{a} - (x_3^{b} + k ))^2)} =  K(x_a, x_b) $.
\end{proof}

\subsubsection{MR-4: Linear Scaling of the Test Features (only for Linear Kernel)}
Let $X_{train}$ be the set of training data and $X_{test}$ be the set of test data. Upon completion of training via SVM, let a particular test instance, $x_{a}$ be classified as class $c$ with a score of $s$. For this MR, we scale only the test instances as follows. Let $x_{b} = 2 * x_{a}$ and $x_{c} = 3 * x_{a}$. Then, the functional distance between $x_{a}$ and $x_{b}$ would be equal to the functional distance between $x_{b}$ and $x_{c}$. Note that the classes of $x_{b}$ and $x_{c}$ need not be $c$. This MR can be used on an already trained model and thus in cases where the training API is not available to the tester. 

\begin{proof}
Since the optimization has already been completed, the optimal values of $\alpha$ have been found through Equation (\ref{svmDual}). Using Equation (\ref{funcDist}), we can reason about the functional distances: 
\[\begin{aligned}
	D(x_b) - D (x_a) & = \sum_{i=1}^{i=m} \alpha_i y_i K(x_b, x_i) + b - \sum_{i=1}^{i=m} \alpha_i y_i K(x_a, x_i) - b \\
	& = 2 \sum_{i=1}^{i=m} \alpha_i y_i {x_a}^T {x_i} - \sum_{i=1}^{i=m} \alpha_i y_i {x_a}^T {x_i} \\
	& \text{Since } x_b = 2 * x_a \text{ and } {(2x)}^Tx = 2 * {(x)}^Tx \\
	& = \sum_{i=1}^{i=m} \alpha_i y_i {x_a}^T {x_i} \\
	\text{Similarly,}\\
	D(x_c) - D (x_b) & = 3 \sum_{i=1}^{i=m} \alpha_i y_i {x_a}^T {x_i} -  2 \sum_{i=1}^{i=m} \alpha_i y_i {x_a}^T {x_i} \\
	& = \sum_{i=1}^{i=m} \alpha_i y_i {x_a}^T {x_i} = D(x_b) - D (x_a)
\end{aligned}
\] \end{proof} 

Each of the four MRs developed, can have multiple variants. For example, in MR-1, the features can be permuted in multiple ways. However, since the outputs from the MRs are expected to be an exact match (within a threshold of $10^{-6}$), we created only one variant for each MR. The variant, however, ensured that every aspect of the data instance was changed. For example, in MR-1, every feature was permuted and in MR-2, the order of every training data instance was changed. Similarly, for MR-3 \& MR-4, a single variant was used.
 
The efficacy of these MRs in terms of the potential to catch implementation bugs is presented in Section \ref{results}.

\subsection{Metamorphic Relations for Application 2: Image Classification using ResNet}
The second application, we have chosen, is a deep learning based image classifier called a Residual Network (ResNet) \cite{he2016deep}. The application classifies color images into $10$ classes. ResNet is considered as a breakthrough advancement in the usage of Convolutional Neural Networks (CNN) for image classification and has been shown to work very well (it won multiple competitions in image classification tasks in 2015 \cite{ILSVRC15} \cite{lin2014microsoft}). The ResNet architecture has also been shown to work well in different contexts \cite{he2016deep} and is now being largely considered as a default choice for image classification \cite{resNetDefaultChoice}.

\begin{figure}
    \centering
    \includegraphics[height=1.5in]{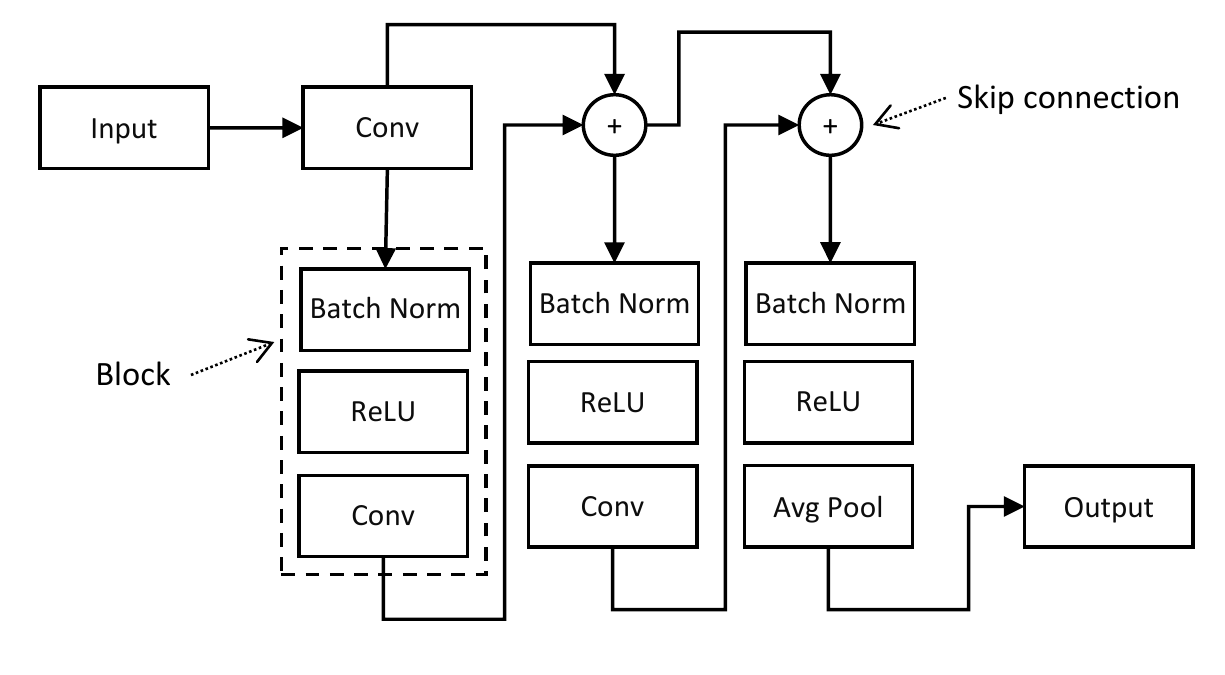}
    \caption{A simplified version of the ResNet Architecture.}
    \label{resnetArch}
\end{figure}

 
We have taken the official implementation of ResNet in TensorFlow \cite{resNetImplementation} (in particular the ResNet-32 configuration). The ResNet architecture comprises of a series of neural network layers and a simplified version is shown in Figure \ref{resnetArch}. The architecture consists of an initial Conv layer followed by a series of `blocks'. Each block consists of a Batch-Norm layer, the activation function of ReLU and another Conv layer. Between each block is a `skip' connection (the skip connection is thought of as the breakthrough idea \cite{resNetImplementation}). Conceptually, the ResNet application is learning a complex, non-linear function, $f(X, W)$, where $X$ is the input data and $W$ is a set of weights. 

\begin{figure}[H]
    \centering
    \includegraphics[width=8cm]{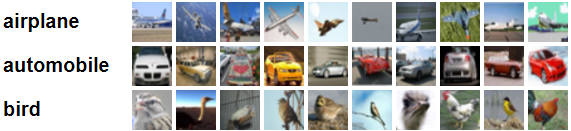}
    \caption{A sample of the training data of 3 classes.}
    \label{cifar10}
\end{figure}

The ResNet application trains and tests over the annotated data instances from the CIFAR-10 corpus \cite{cifar10}. Each data instance is a 32 by 32 pixel color image - i.e. each instance has 32 x 32 x 3 features (numbers) and has been categorized into 10 mutually exclusive classes. A few data instances are visualized in Figure \ref{cifar10}. For a given set of test instances, the ResNet application predicts the classes and outputs the test classification accuracy and the `test loss'. The test loss (which is computed as the `cross-entropy' between the actual class and the predicted class) is a measure of the goodness of the classification. This is analogous to the `score' in SVM. 

We have developed the following metamorphic relations for the ResNet application.

\begin{enumerate}
    \item MR-1: Permutation of input channels (i.e. RGB channels) for training and test data
    \item MR-2: Permutation of the Convolution operation order for training and test data
    \item MR-3: Normalizing the test data
		\item MR-4: Scaling the test data by a constant
\end{enumerate}

\subsubsection{MR-1: Permutation of Input Channels (i.e. RGB Channels)}
The RGB channels of an image hold the pixel values for the `red', `green' and `blue' colors. This data is typically represented in a fixed order of R, G and B. For example, in the CIFAR-10 data, the first 1024 bytes are the values of the R channel, the next 1024 bytes are the values of the G channel, finally followed by 1024 bytes of the B channel. In total, the 3072 bytes forms the data for one image. 

Let $X_{train}$ denote the training dataset and $X_{test}$ denote the set of test data. Given this original data, let the ResNet application complete training and output a test loss and accuracy. The MR claims, if the RGB channels of both $X_{train}$ and $X_{test}$ are permuted (for example, the permutation can be: the first 1024 bytes of the data would contain the B channel, followed by 1024 bytes of the G channel, ending with 1024 bytes of the R channel), a correctly implemented ResNet application should still be able to learn the same (or very similar) set of weights and therefore output the same (or very similar) test loss and accuracy. 

This MR is similar, in principle, with the MR-1 of the SVM application (permutation of features). However, in the case of ResNet, we cannot do any arbitrary permutation - for example, we cannot permute pixel 1 with pixel 20 and vice-versa (as was done in the case of SVM). This is because, the CNN layer is built to exploit the `locality of pixel dependencies' of an image - where groups of pixels close to each other tend to carry similar semantics \cite{krizhevsky2012imagenet}. Thus, the MR-1 (and MR-2 as we will see later) make specific permutations such that this locality property is not violated. 

\begin{figure}
	\subfloat[Original training data (RGB)]{%
	\includegraphics[height=1.5cm]{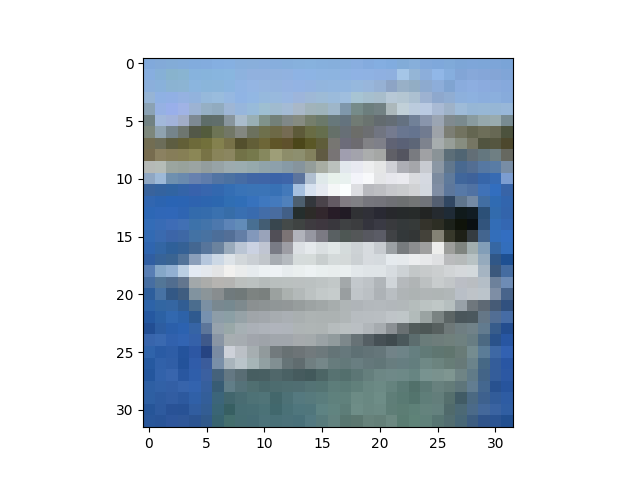}}%
	\hspace{10pt}%
	\subfloat[BGR]{%
	\includegraphics[height=1.5cm]{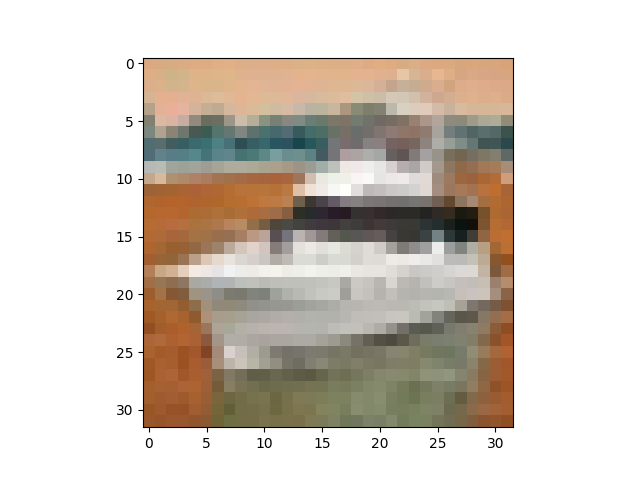}}%
	\hspace{10pt}%
	\subfloat[BRG]{%
	\includegraphics[height=1.5cm]{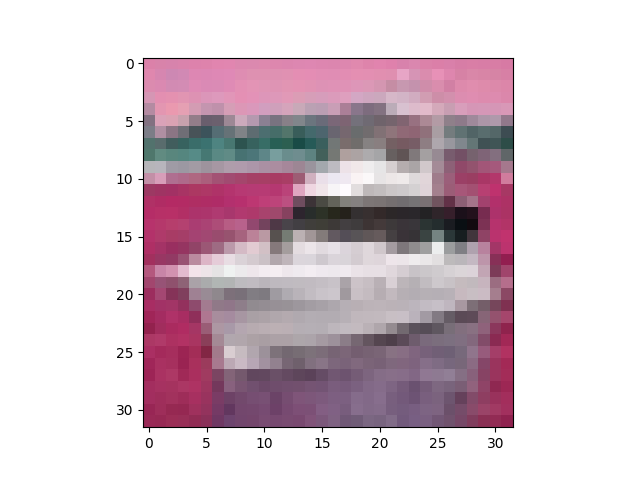}}%
	\hspace{10pt}%
	\subfloat[GBR]{%
	\includegraphics[height=1.5cm]{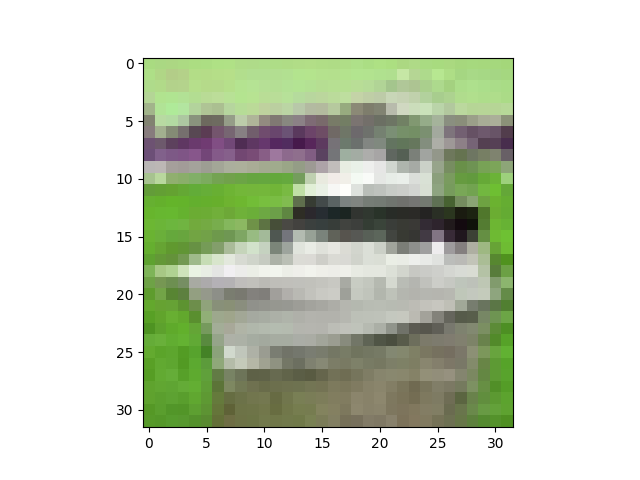}}%
	\hspace{10pt}%
	\subfloat[GRB]{%
	\includegraphics[height=1.5cm]{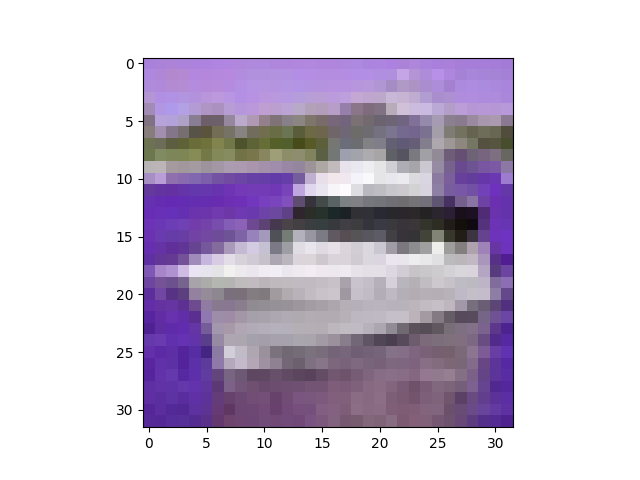}}%
	\hspace{10pt}%
	\subfloat[RBG]{%
	\includegraphics[height=1.5cm]{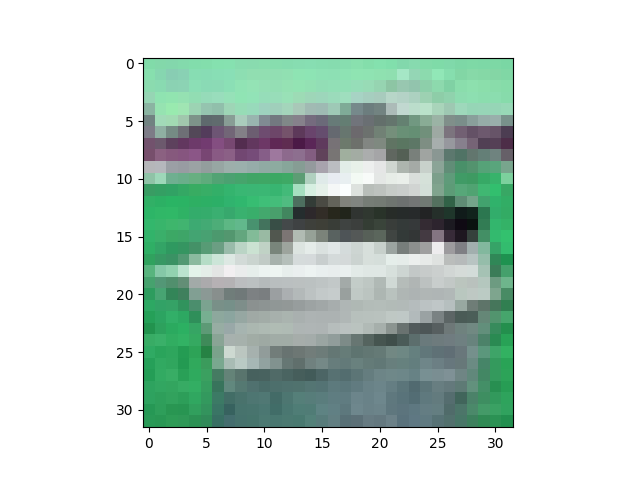}}%
	\caption{Permutation of RGB channels for one instance of training data. The test data is permuted in similar fashion. The results should be very similar whether the ResNet application is trained \& tested on the original or permuted data.}
	\label{mr1_resnet}
\end{figure}

The RGB channels can be permuted in $6$ ways, one of which is the original data (RGB channel in order). We use all 6 variants for this metamorphic relation. The variants are visualized in Figure \ref{mr1_resnet}.

\begin{proof}[Reasoning]
The validity of the MR can be explained in two steps. In the first, we will show that the core property of a CNN (i.e. the `locality property') is not violated by the RGB permutation. This is done by showing that the set of weights which were found while training on the original data is a valid solution on the permuted data. However, we cannot claim that this same set of weights will be found by the optimization method (unlike in the case of SVM). Thus, in the second step, we will show through some empirical results (both ours and those from existing literature) that in practice, a set of weights very close to the original is found.

The first CNN layer of ResNet architecture (refer Figure \ref{resnetArch}) has a weight for each of the three channels. Consider the first pixel of an image, which can be represented as $I = [i^r, i^g, i^b]$. Similarly, let the first weight of the CNN layer be represented as $W = [w^r, w^g, w^b]$. Then the convolution operation gives the result: $CONV(I, W) = i^r w^r + i^g w^g +i^b w^b $.

Now, let the permuted input and weight be: $I^p = [i^b, i^g, i^r] \& W^p = [w^b, w^g, w^r]$. Then, $CONV(I^{p}, W^{p}) = i^b w^b + i^g w^g +  i^r w^r = CONV(I, W)$. Thus, the output from the first layer would be the same as before, and this implies all subsequent outputs (including the final classification) would be the same as before. The maintenance of the locality property can also be seen in Figure \ref{mr1_resnet} where the images are still semantically recognizable.

However, it isn't clear whether the optimization method used in the ResNet application (which is based on gradient descent) can indeed arrive at the set of weights $W^p$. At the start of training, the weights of the model (the parameters) are initialized to some random values. The weights are then changed as per the gradient of the loss ($L$) w.r.t the weights ($\frac{\partial}{\partial w}L$). Changing the input data from RGB to BGR makes no difference to the initialization of the weights, however, the gradient changes with the change in input. This means that the weights start from the same point in the weight space for RGB and BGR, but start moving in different directions   and can converge at different points (i.e. the error surfaces are different for RGB \& BGR). 

The impact of change of channels can be conceptualized in another way. Instead of thinking of changing error surfaces, we can instead consider the impact of changing RGB to BGR as though the initialization of the weights for BGR is in a different point in the error surface of the RGB data. Thus, changing channels essentially means changing the initialization points and this can lead to different convergence. However recent empirical evidence \cite{Goodfellow2014} \cite{choromanska2015loss} \cite{dauphin2014identifying} suggests that practically, these different convergence points are very close to each other in terms of overall loss. Thus, it is expected that irrespective of the change in  input from RGB to BGR, the weights to which they converge, must be very close to each other in terms of the overall loss. \end{proof}

\begin{figure*}
	\subfloat[MR-1 on ResNet code with CIFAR10 data]{
	\includegraphics[width=0.23\textwidth]{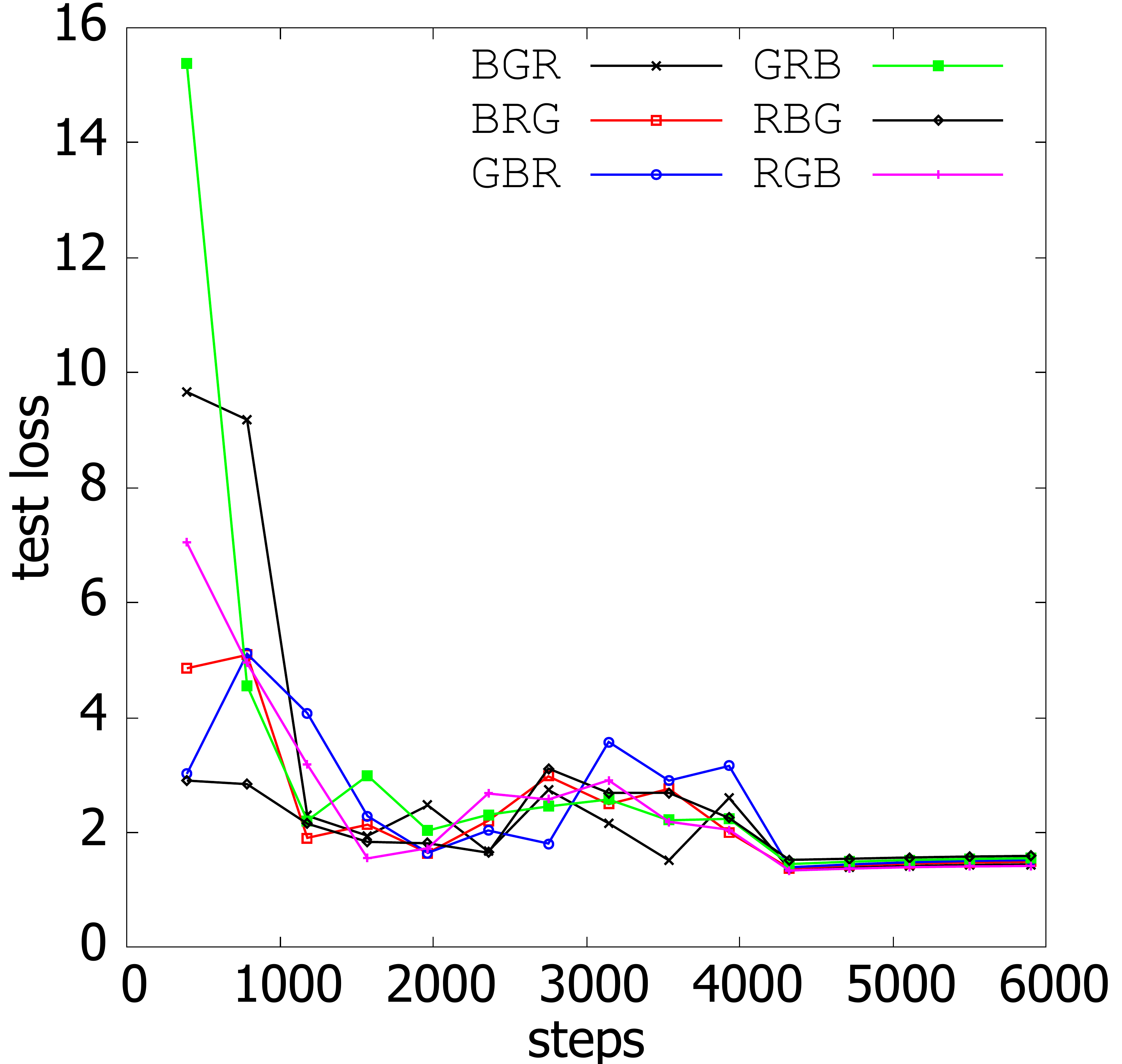}}%
	\hspace{8pt}%
	\subfloat[MR-1 on ResNet code with SVHN data]{
	\includegraphics[width=0.23\textwidth]{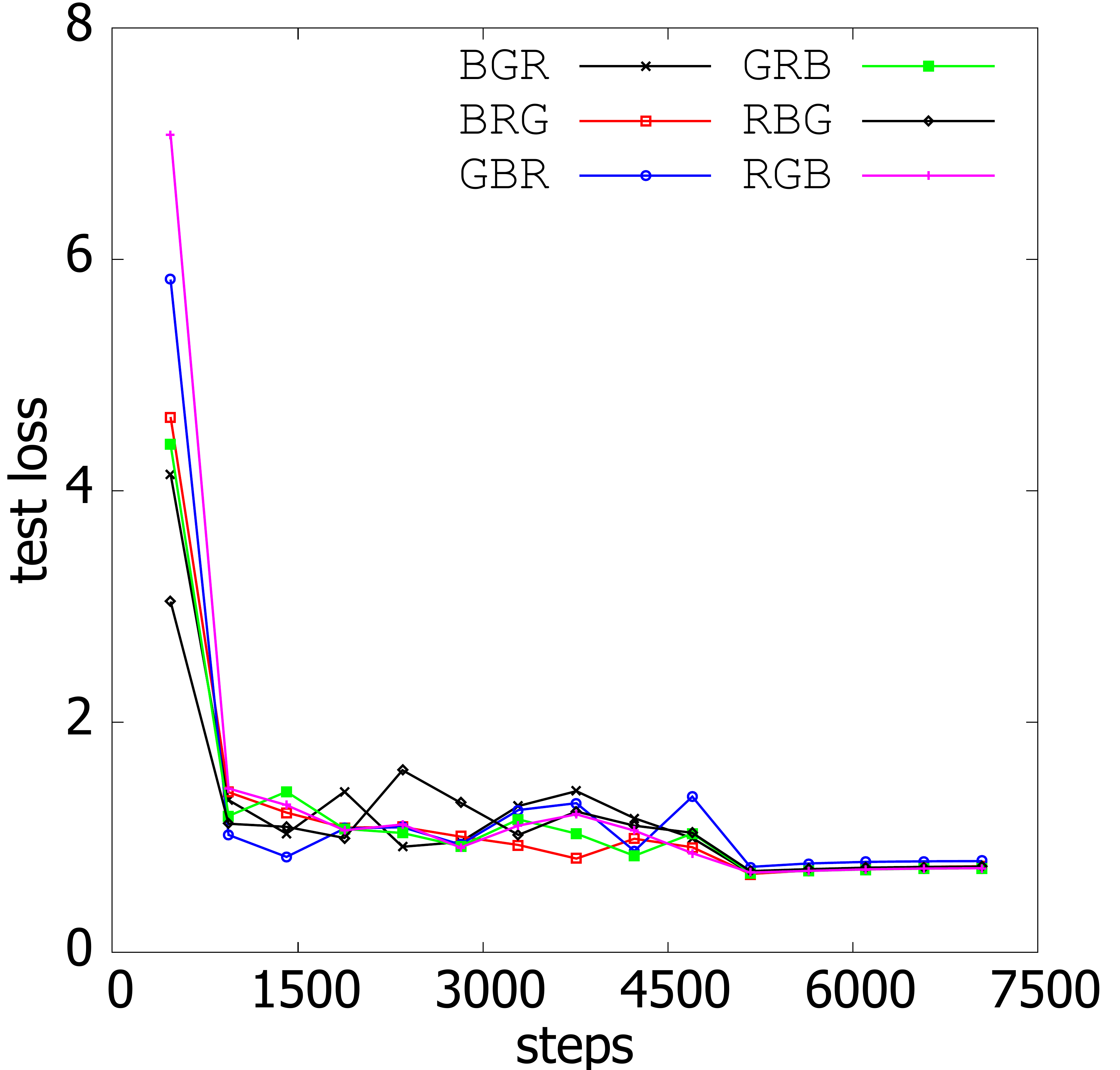}}%
	\hspace{8pt}%
	\subfloat[MR-1 on AlexNet code with CIFAR10 data]{
	\includegraphics[width=0.23\textwidth]{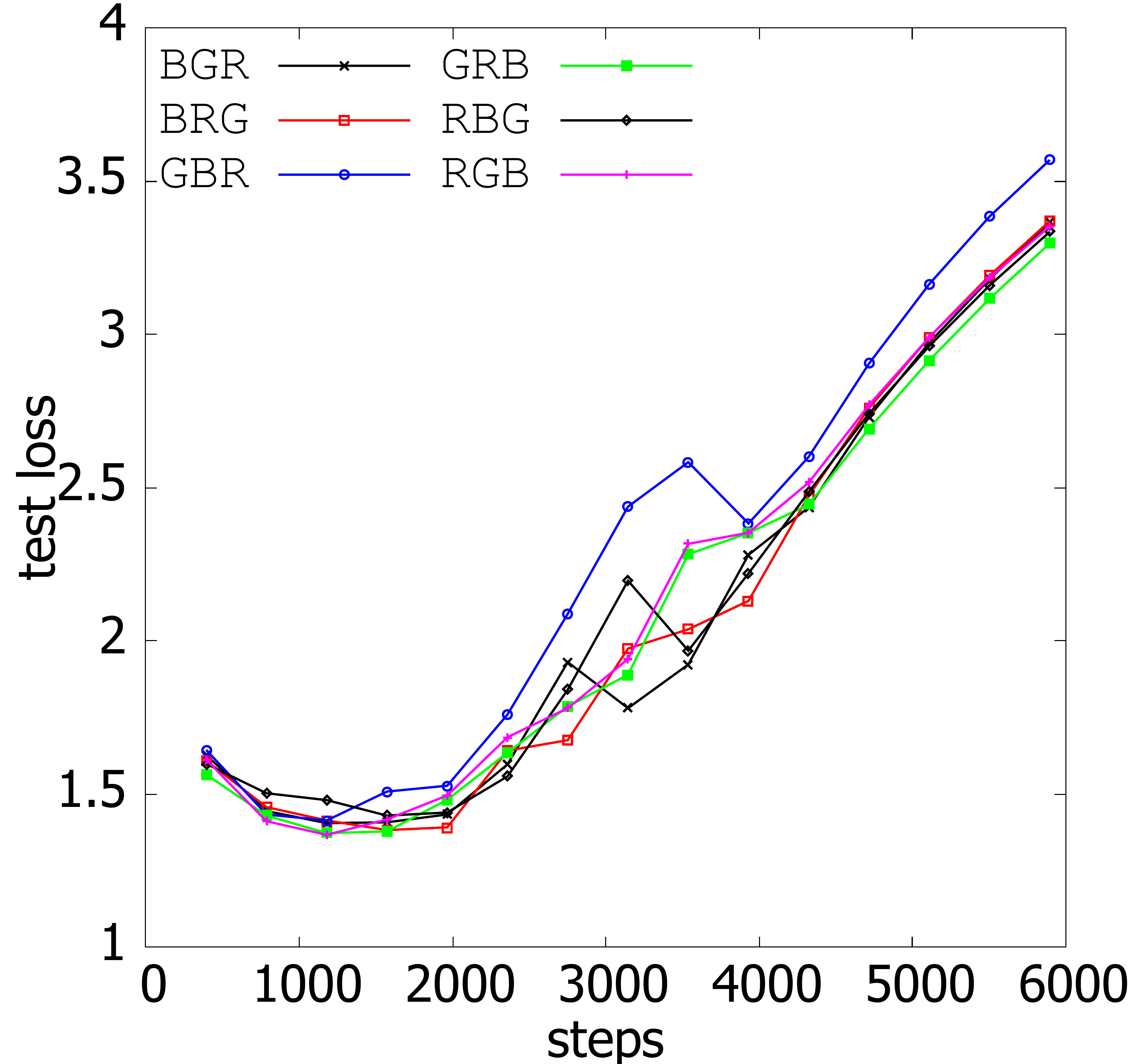}}%
	\hspace{8pt}%
	\subfloat[MR-1 on VGGNet code with CIFAR10 data]{
	\includegraphics[width=0.23\textwidth]{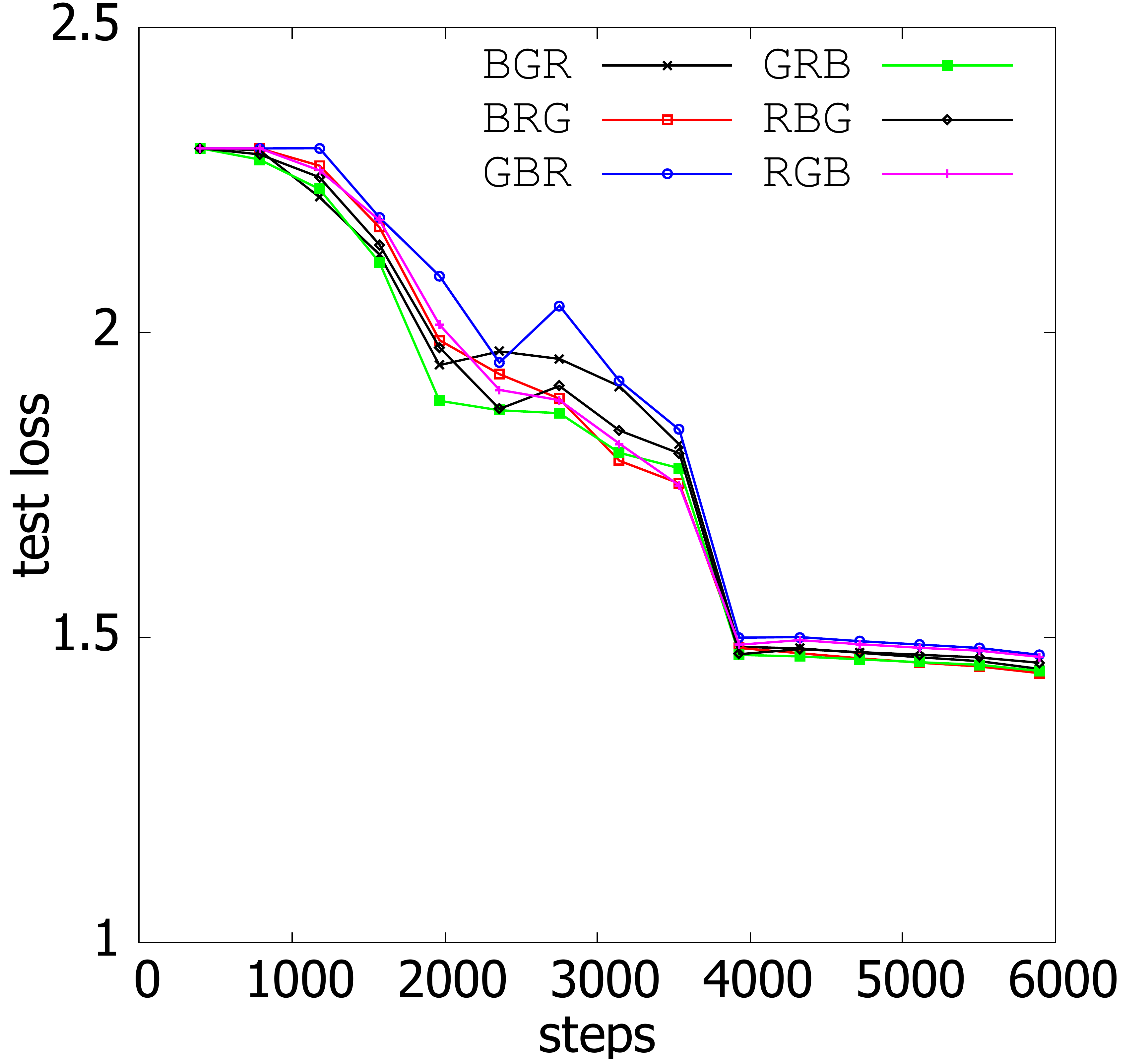}}%
\\%
	\subfloat[MR-2 on ResNet code with CIFAR10 data]{
	\includegraphics[width=4cm]{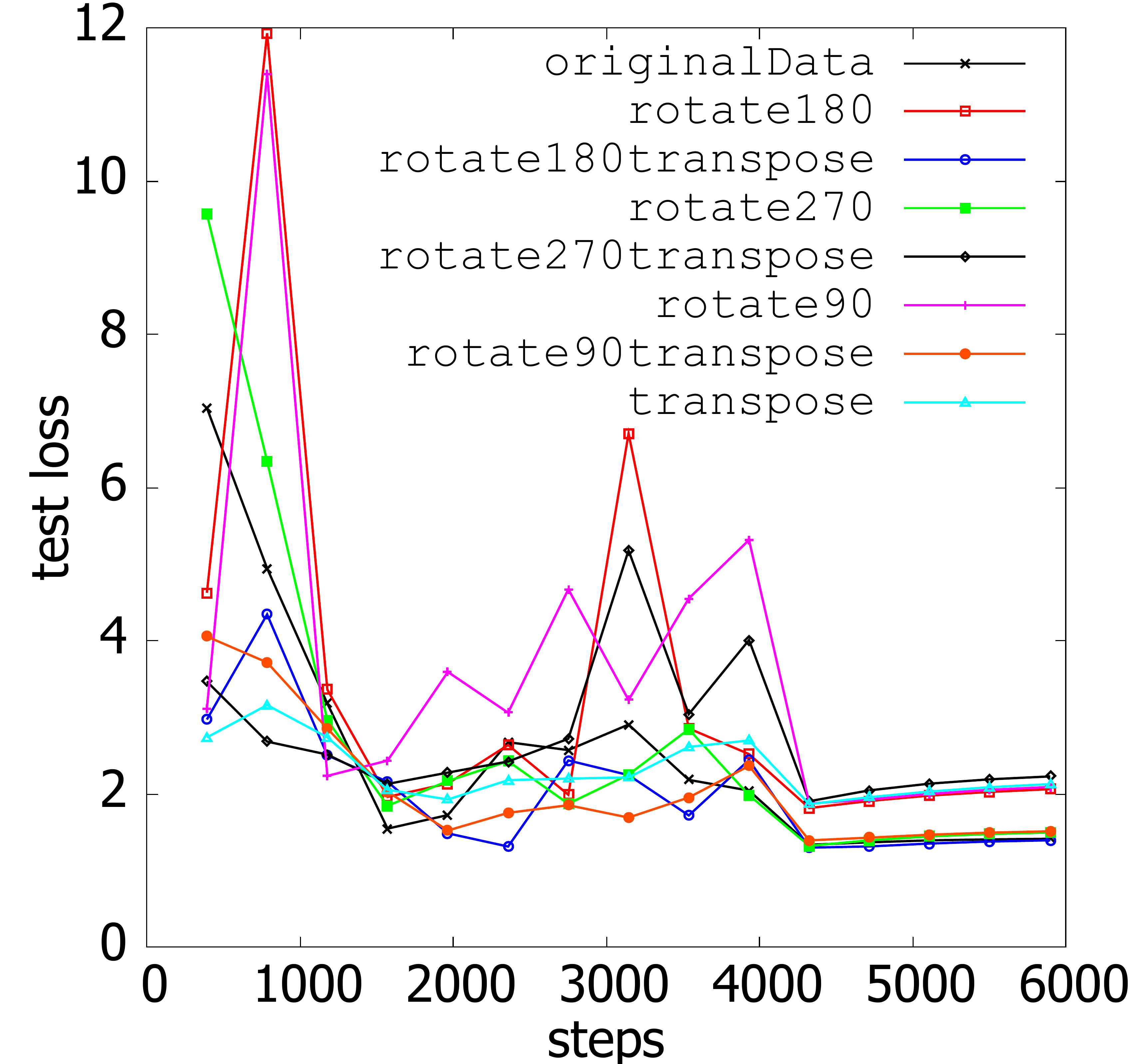}}%
	\hspace{8pt}%
	\subfloat[MR-2 on ResNet code with SVHN data]{
	\includegraphics[width=4cm]{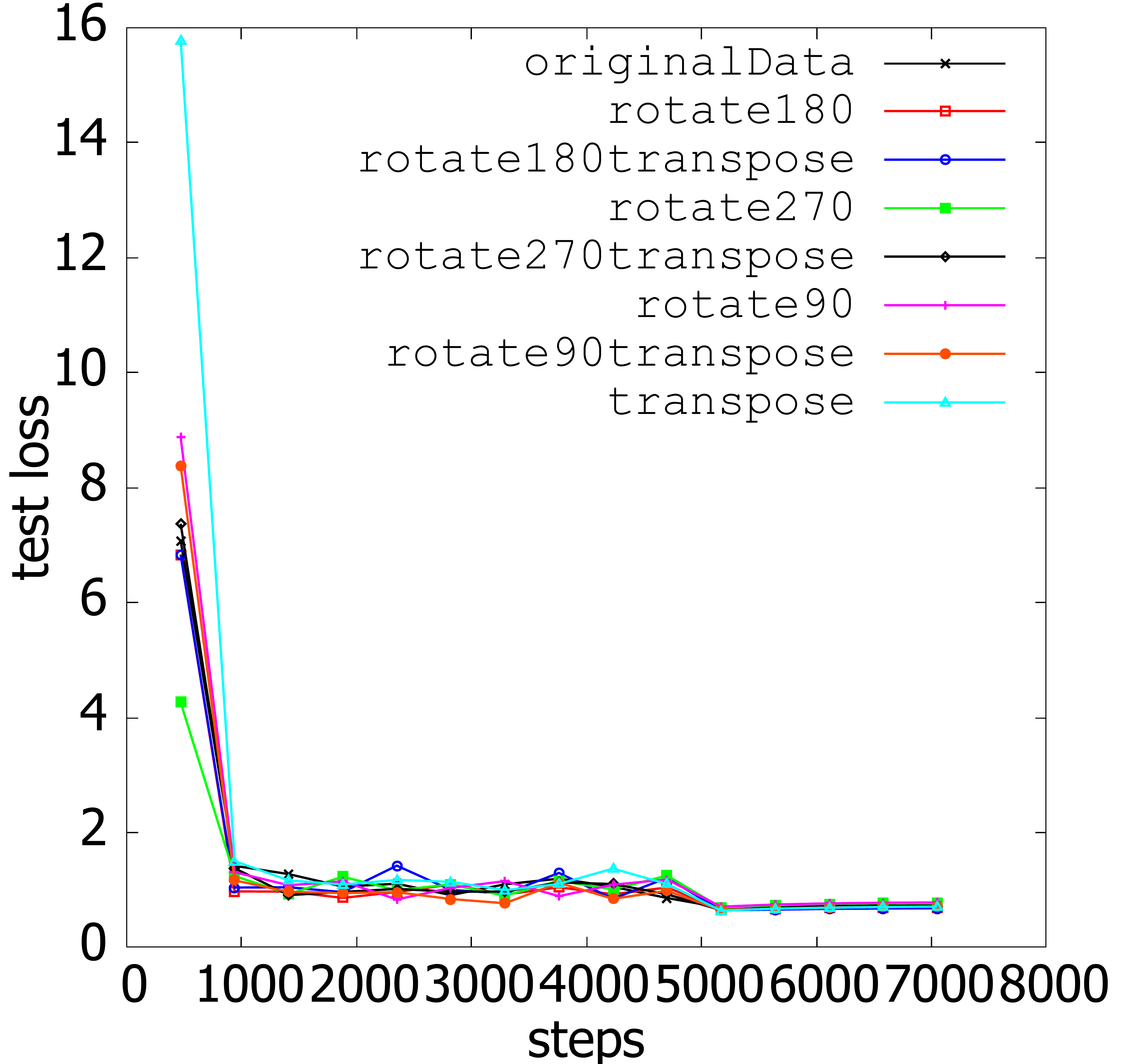}}%
	\hspace{8pt}%
	\subfloat[MR-2 on AlexNet code with CIFAR10 data]{
	\includegraphics[width=4cm]{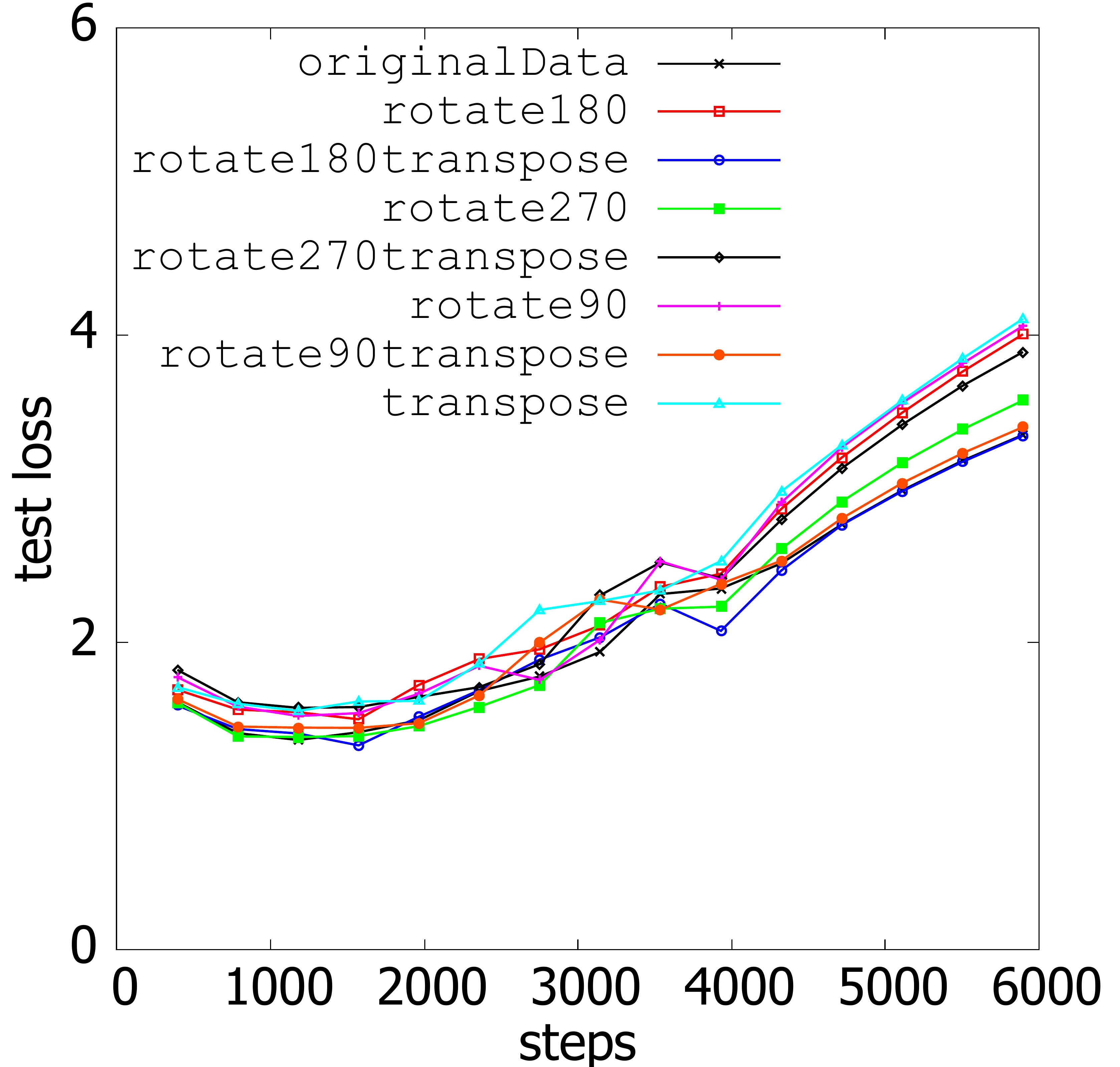}}%
	\hspace{8pt}%
	\subfloat[MR-2 on VGGNet code with CIFAR10 data]{
	\includegraphics[width=4cm]{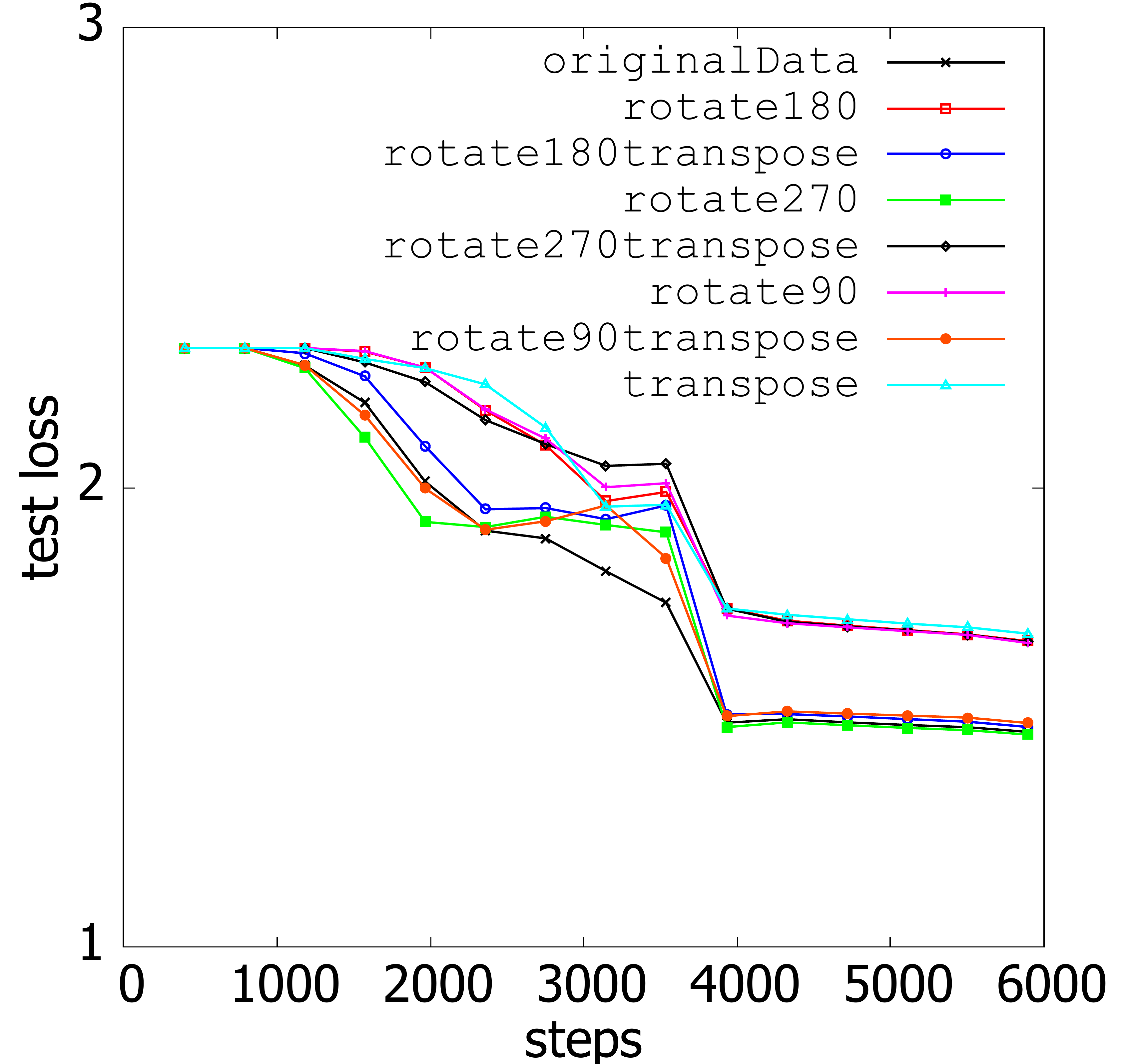}}%
%
	\caption{Test loss from original code (non-buggy) due to changes in RGB order per MR-1 (top four graphs) and permuting input per MR-2 (bottom four graphs) for different datasets \& architectures. Please see Appendix A \cite{ourAppendix} for complete results.}
	\label{validationLoss_originalCode}
\end{figure*}

\begin{proof}[Empirical Evidence] To give further credence to this MR, we conducted a set of empirical tests. We selected three additional datasets and three different architectures (shown in Table \ref{testMR1different}) to validate the MR. The variation in loss (on the test data) for the experiments as the training progresses is shown in Figure \ref{validationLoss_originalCode}. Due to space considerations, we show the results on two datasets and two architectures. Complete data is available in Appendix A \cite{ourAppendix}. Notice that there is little variation in the curves for the RGB variants and this is also measured as the maximum of the standard deviation ($\sigma_{max}$) in Table \ref{testMR1different}.

The combination of the tests showed that permuting the order of the RGB channels does not significantly change the results. From Table \ref{testMR1different}, we can also observe that this property holds for different architectures and different datasets.
\end{proof}

\begin{table}[H]
\caption{Experiments conducted to validate MR-1 \& MR-2}
	\begin{tabu} {|X[-1.2,c,m]|X[-1.3,c,m]|X[c,m]|X[c,m]|}
		\hline
		Deep Learning Architecture & Dataset & $\sigma_{max}$ in test loss due to MR-1 (permute RGB) & $\sigma_{max}$ in test loss due to MR-2 (permute CONV order)\\
		\hline
		ResNet & Cifar10 & 4.8 & 3.6 \\
		\hline
		ResNet & SVHN \cite{netzer2011reading} &1.4 & 3.3 \\
		\hline
		ResNet & Kaggle Fruits \cite{kaggleFruits} & 0.8& 2.1  \\
		\hline
		ResNet & Kaggle digits \cite{kaggleHandWritten} &1.1 & 0.9 \\
		\hline
		AlexNet \cite{krizhevsky2012imagenet} & Cifar10 &0.3 & 0.3\\
		\hline
		VGGNet \cite{simonyan2014very} & Cifar10 & 0.1& 0.1\\
		\hline
		NIN \cite{lin2013network} & Cifar10 &0.2 & 0.2\\
		\hline
	\end{tabu}
\label{testMR1different}
\end{table}
 
\subsubsection{MR-2: Permutation of the Convolution Operation Order for Training and Test Data}
This MR, again, is similar to the case of permutation of features for SVM (Section \ref{svmMR1}). However, we need to explicitly maintain the locality property of the permuted features. Our metamorphic relation, is to permute the input pixels such that the local neighborhood is maintained - i.e  pixels which were neighbors in the original image, continue to be neighbors in the permuted image. To explain this, consider an image $I$ with pixel values as shown in Figure \ref{mr2_convOrder_pixel}. $I$ contains 3 x 3 pixels and $I^{T}$ represents a permutation of the image such that the neighbors of every pixel are maintained. The permutation shown is a matrix transpose.  

\begin{figure}[H]
\[
I = 
\begin{bmatrix}
    1  & 2 & 3  \\
    4  & 5 & 6    \\
    7  & 8 & 9 
\end{bmatrix} 
\ \ \ 
I^{T} = 
\begin{bmatrix}
    1  & 4 & 7  \\
    2  & 5 & 8    \\
    3  & 6 & 9 
\end{bmatrix}
\]
\caption{An example image $I$ \& it's permutation $I^{T}$ which maintains the neighbors of every pixel}
\label{mr2_convOrder_pixel}
\end{figure}

The MR is as follows. Let $X_{train}$ and $X_{test}$ be the training and test data respectively. After training the ResNet application with this data, let a instance of the test data $x_{test}^i$ be classified as class $c$ with a loss of $s$. If we permute the training and test data such that the neighborhood property is maintained and retrain the ResNet application on the permuted data, the (permuted) test instance will continue to be classified as class $c$ with a loss of $s$.

\begin{figure}[H]
	\subfloat[Original training data]{%
	\includegraphics[height=1.5cm]{figures/ship1-Orig.png}}%
	\hspace{3pt}%
	\subfloat[Matrix Transform of original]{%
	\includegraphics[height=1.5cm]{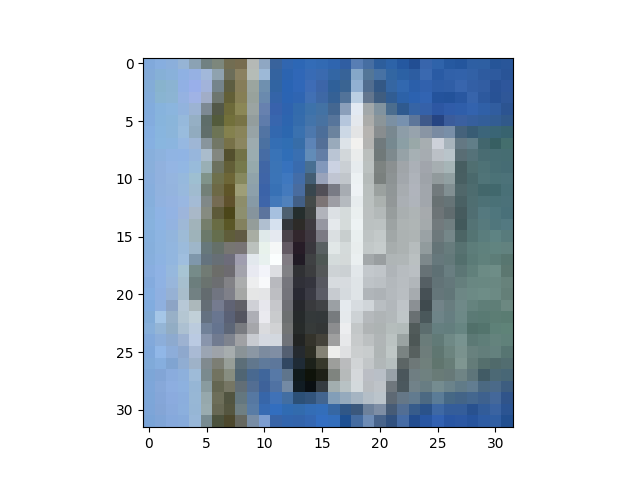}}%
	\hspace{5pt}%
	\subfloat[$90^{o}$ rotation of original]{%
	\includegraphics[height=1.5cm]{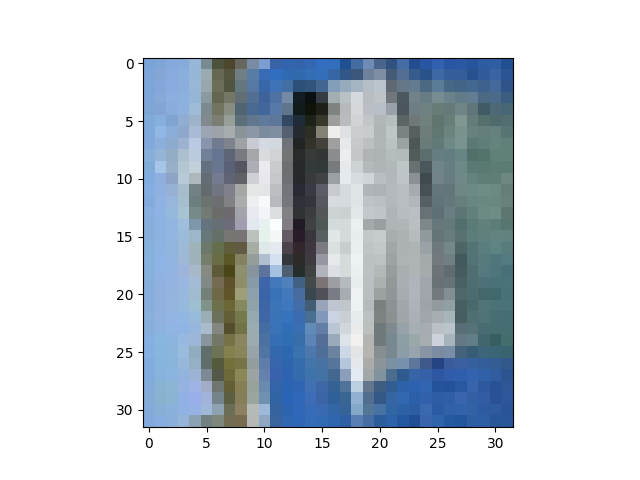}}%
	\hspace{5pt}%
	\subfloat[Transform of $90^{o}$ rotation]{%
	\includegraphics[height=1.5cm]{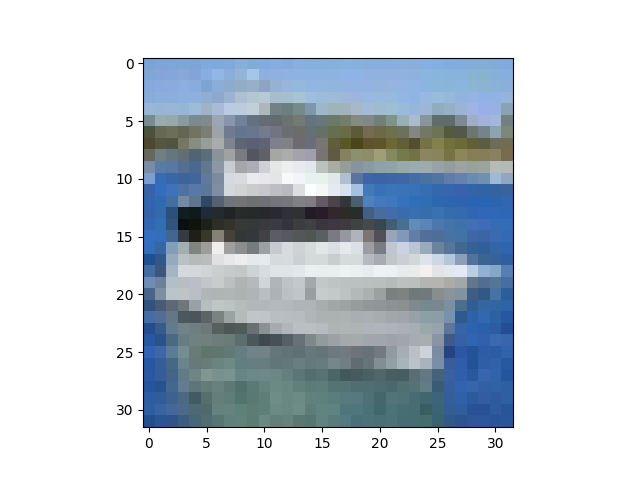}}%
	\\
	\subfloat[$180^{o}$ rotation]{%
	\includegraphics[height=1.5cm]{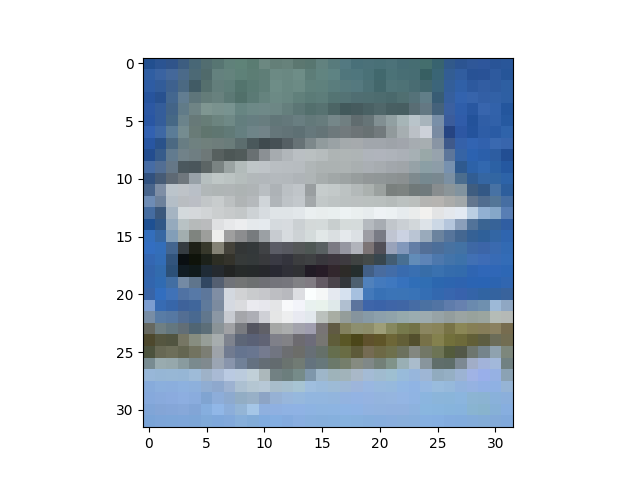}}%
	\hspace{3pt}%
	\subfloat[Transform of $180^{o}$ rotation]{%
	\includegraphics[height=1.5cm]{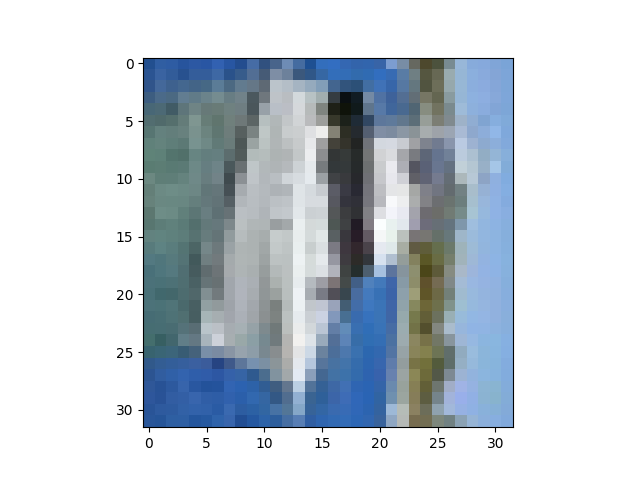}}%
	\hspace{5pt}%
	\subfloat[$270^{o}$ rotation]{%
	\includegraphics[height=1.5cm]{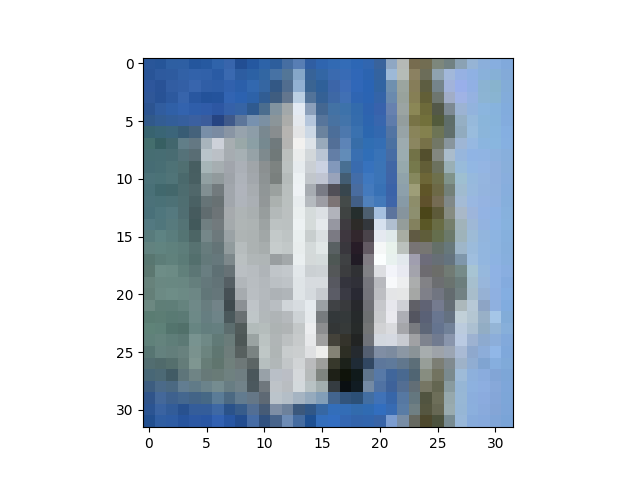}}%
	\hspace{5pt}%
	\subfloat[Transform of $270^{o}$]{%
	\includegraphics[height=1.5cm]{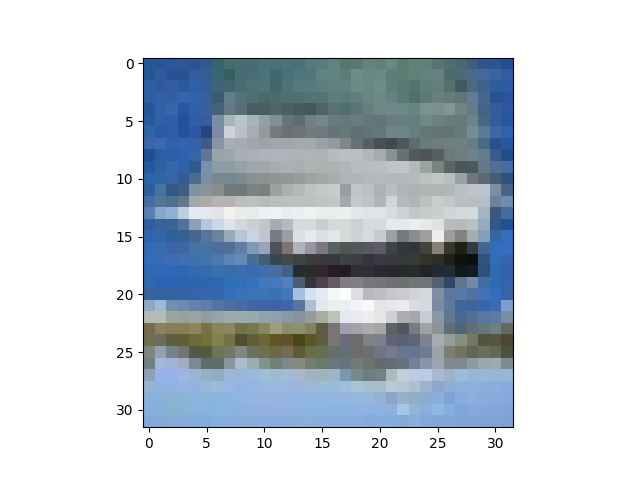}}%
	\caption{Permutation of CONV order for one instance of training data. The test data is permuted in similar fashion}
	\label{mr2_visuals}
\end{figure}

We have identified $7$ permutations that maintain the neighborhood property - matrix transpose, $90^o$ rotation, $180^o$ rotation, $270^o$ rotation \& matrix transpose of each rotation. These $7$ transformations are visualized in Figure \ref{mr2_visuals}. We use these 7 variants along with the original image for MR-2. 

\begin{proof}[Reasoning]
The reasoning is similar to that of MR-1. Let the weight matrix of the first CONV layer (refer Figure \ref{resnetArch}) after training be $W$. Let $I$ be one test image. If we permute $I$ as $I^{T}$, and permute $W$ as $W^{T}$, then the output from the first CONV layer will be the same as before, except in permuted order. For example, consider an image $I$ and weight $W$ shown in Figure \ref{convOrderExample}. We can see that $CONV(I, W) = {CONV(I^{T}, W^{T})}^{T}$. Thus, the output at each layer is a transpose of the earlier output. This property holds for all the layers in ResNet (including ReLU, BatchNorm and the skip connections). Thus, at the final layer, we would get the same output as before, but in transposed form. The weights at this final layer, again when transposed, will give the same classification and loss as before. 

\begin{figure}[!htb]
	\[
		I = 
			\begin{bmatrix}
				1  & 2 & 3  \\
				1  & 1 & 2    \\
				2  & 0 & 1 
			\end{bmatrix} 
		\ \ 
		,\ W =
			\begin{bmatrix}
				1  & 0   \\
				2  & 3    
			\end{bmatrix}
		\ \ ;\ 
		CONV(I, W) = 
			\begin{bmatrix}
				6  & 10   \\
				5  & 4     
			\end{bmatrix}
   \]
	
	\[
		I^{T} = 
			\begin{bmatrix}
				1  & 1 & 2  \\
				2  & 1 & 0    \\
				3  & 2 & 1 
			\end{bmatrix} 
		\ \ 
		,\ W^{T} =
			\begin{bmatrix}
				1  & 2   \\
				0  & 3    
			\end{bmatrix}
		\ \ ;\ 
		CONV(I^{T}, W^{T}) = 
			\begin{bmatrix}
				6  & 5   \\
				10  & 4     
			\end{bmatrix}
   \]
\caption{Example showing the impact of MR-2. }
\label{convOrderExample}
\end{figure}

However, as in case of MR-1, we cannot claim that the optimization method will find the transposed set of weights. Nevertheless, the transposition of the input can be conceptualized as a different point in initialization and we can fallback on the empirical evidence which has shown that small changes in initialization do not significantly change the final convergence qualities.
\end{proof}

\begin{proof}[Empirical Evidence] Similar to MR-1, we conducted empirical tests to validate MR-2. We took three different datasets and three different deep learning architectures. MR-2 (with 7 variants in the permutation of the inputs) was compared with the original data. Figure \ref{validationLoss_originalCode}(e-h) shows the results. As before, we found the deviation of test loss after 150 epochs to be small. The deviation is captured as $\sigma_{max}$ in Table \ref{testMR1different}.
\end{proof}

\textbf{A note on MR-1 \& MR-2}: It is a common practice to perform data-augmentation in ML applications. Data-augmentation includes cases such as mirroring images and some of the transformations in Figure \ref{mr2_visuals} might look like cases of data-augmentation. However, our MRs are clearly distinct from data-augmentation. The intuition behind data-augmentation is that of `validation' - i.e. a mirror-image is a naturally occurring phenomenon. Further, such mirror-images are computed and added to the training data where both the original and the mirror-image exist in the training data. 

However, the transformation in our MRs is a case of `verification'. Some of the transformations may not naturally occur (many cases of RGB transformation look wrong to the naked eye). The MRs claim that the classification of an original image and the transformed image should lie in the same class (even if the class is wrong). Further, our transformations do not add to the training data - i.e. the transformed and the original image never co-exist in the training data.

\subsubsection{MR-3: Normalizing the Test Data}
A standard practice in machine learning applications is to normalize the training and test data. Normalization makes the data with a zero mean \& unit variance \cite{resNetDefaultChoice} and leads to faster convergence. 

\begin{figure}[H]
	\subfloat[Original test image]{%
	\includegraphics[height=1.5cm]{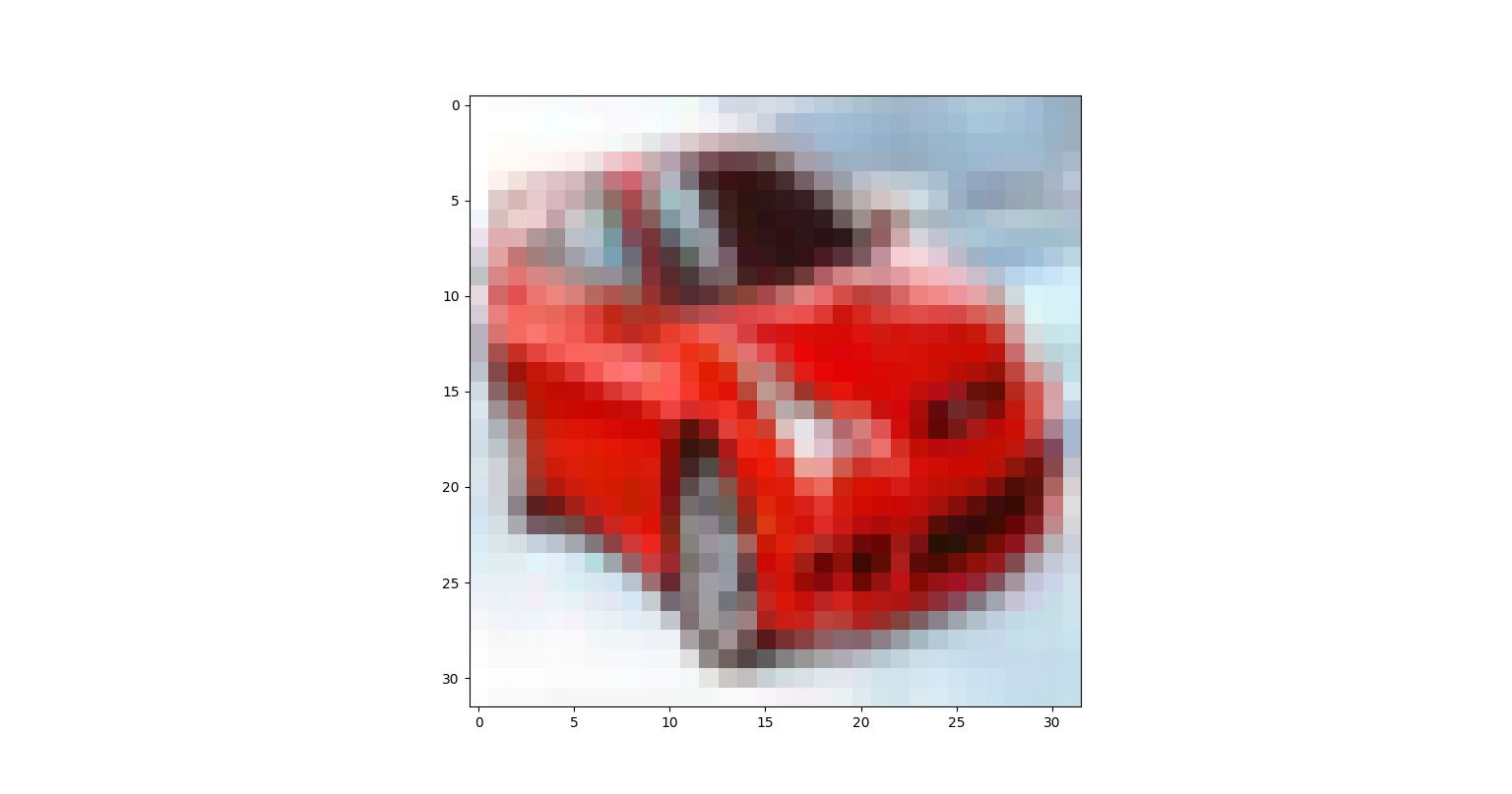}}%
	\subfloat[Normalized test image]{%
	\includegraphics[height=1.5cm]{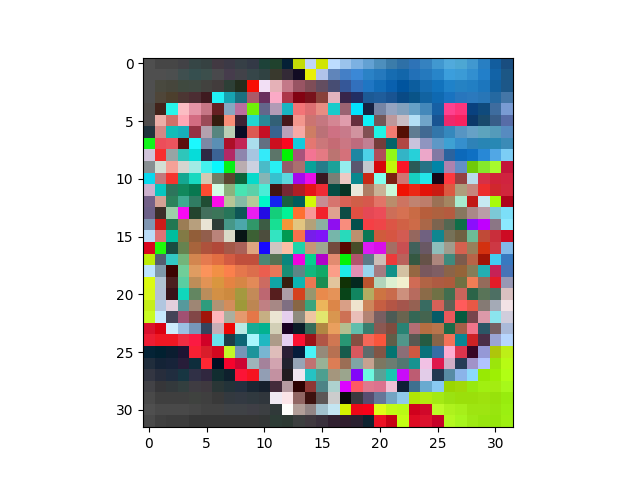}}%
\caption{Visualisation of MR-3. The class \& loss should be the same when tested on the original or normalized images.}
\label{mr3visuals}
\end{figure}

Let $X_{train}$ be the training data and $x_{test}^i$ be one instance of the test data. Let the ResNet application complete its training on $X_{train}$. Let the classification of $x_{test}^i$ be $c$ with a loss of $s$. MR-3 specifies that if we normalize $x_{test}^i$ and feed this as the input the results would be exactly the same. This is visualized in Figure \ref{mr3visuals}. Note that this MR needs only the test data to be normalized and therefore can be used to test already trained ML models.

\begin{proof}
There are two forms of normalization that can be performed. In one, the normalization is done on the entire training data (or a minibatch). In the second form (as is done in ResNet), the normalization is done on individual data points. The proof below is for the latter version.

Let $x$ be a test image (an array of 32 x 32 x 3 numbers). The ResNet application first normalizes this data point (as per Equation (\ref{meanNorm1})) before trying to classify it. 
\begin{equation}
	x' = \frac{x-\mu(x)}{\sigma(x)}
\label{meanNorm1}
\end{equation}

Now, if we normalize $x$ before providing it to the ResNet application, the second normalization that is done inside the ResNet application will not make any difference (as shown in Equation (\ref{meanNorm2})). Thus, the results should be the same whether we supply the original test input or the normalized test input.
\begin{equation}
	x'' = \frac{x'-\mu(x')}{\sigma(x')} \\
	= x'\ \ since\ \mu(x') = 0\ \&\ \sigma(x') = 1
\label{meanNorm2}
\end{equation}
\end{proof}

\subsubsection{MR-4: Scaling the Test Data by a Constant}
Let the ResNet application complete the training on the original training data $X_{train}$. Let $x_{test}^{i}$ be an instance of the test data which is classified into class $c$ with a loss of $s$. MR-4 specifies that if every feature of the test instance $x_{test}^{i}$ is multiplied by a positive constant, $k$ and re-tested, the classification will continue to be $c$ with loss $s$. 

\begin{figure}[!htb]
	\subfloat[Original test image]{%
	\includegraphics[height=1.5cm]{figures/automobile10-orig.png}}%
	\subfloat[$k=\frac{1}{2}$]{%
	\includegraphics[height=1.5cm]{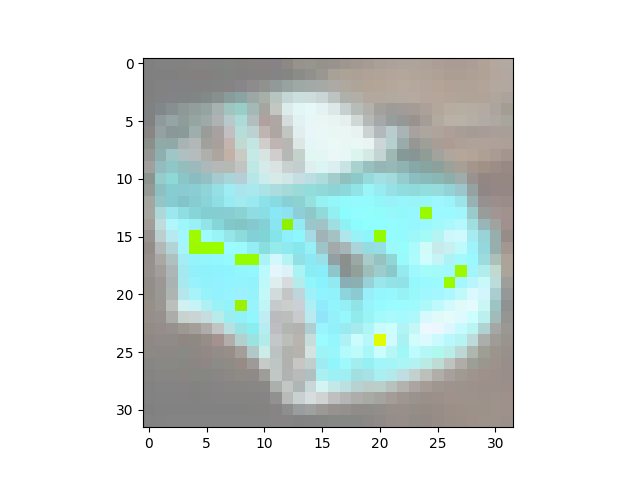}}%
	\subfloat[$k=2$]{%
	\includegraphics[height=1.5cm]{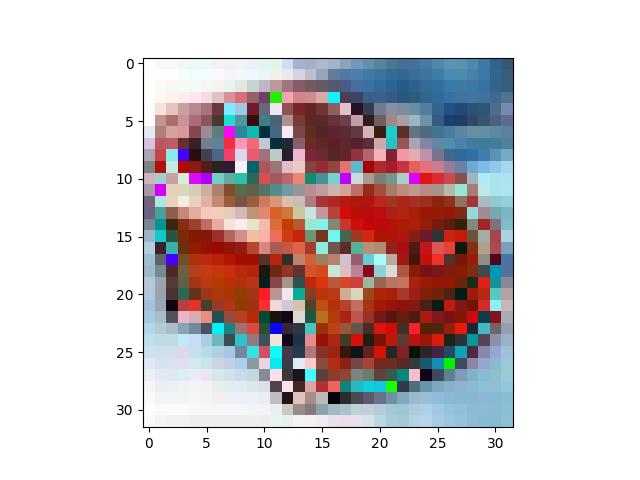}}%
	\subfloat[$k=29$]{%
	\includegraphics[height=1.5cm]{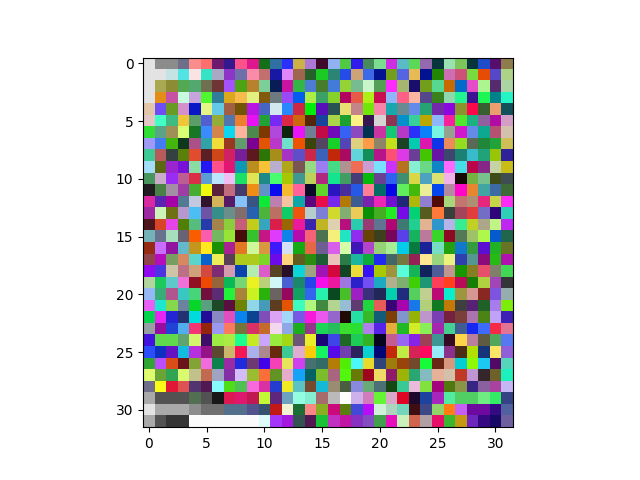}}%
\caption{Visualization of MR-4 (each feature is multiplied by a constant $k$). The class \& loss should be the same when tested on the original or scaled images.}
\label{mr4Visuals}
\end{figure}

\begin{proof}
Like MR-3, MR-4 is a consequence of the type of normalization used in ResNet. When every feature of the input image is multiplied by a constant, we have:

\begin{equation}
	\begin{split}
		x' = \frac{kx-\mu(kx)}{\sigma(kx)} = \frac{k(x-\mu(x))}{k(\sigma(x))} = \frac{x-\mu(x)}{\sigma(x)} \\
		\ \ since \ \mu(kx) = k\mu(x); \sigma(kx) = k\sigma(x) 
		\label{mr4Equation}
	\end{split}
\end{equation}
\end{proof}

Thus, the results should be same irrespective of the constant $k$. Like MR-3, this MR is applied only on the test data and therefore can be used to test already trained ML models.

MR-3 \& MR-4 do not need an empirical validation as they are exact - i.e. any small deviations (we check for a variance $\geq 0.1$) in the outputs are sufficient to indicate the presence of a bug.

We now present the empirical results to validate the efficacy of the MRs developed.

\section{Empirical Results} \label{results}
The efficacy of the MRs developed for the two ML applications was tested by purposefully introducing implementation bugs through the technique of Mutation Testing \cite{jia2011analysis}. Mutation testing systematically changes lines of code in the original source file and generates multiple new source code files. Each file will have exactly one bug introduced. Such a file with a bug is called a mutant. The principle underlying Mutation Testing is that the mutants generated represent the errors programmers typically make \cite{jia2011analysis}. This approach of using Mutation Testing to validate the efficacy of Metamorphic Testing has been used in the past as well\cite{xie2011testing}\cite{liu2014effectively}.

We used the MutPy tool \cite{mutpy}, from the Python Software Foundation, to generate the mutants. Each mutant was executed as per the MRs developed. If any of the results from the mutants did not adhere with what was specified by the MRs, we would term the mutant as `killed' - i.e. the presence of a bug has been identified.

\subsection{Results from Metamorphic Testing of Application 1}
\subsubsection{Creating the Mutants}
The SVM application supports the linear kernel and the RBF (a non-linear) kernel. Using MutPy tool \cite{mutpy}, mutants for both variants were created. In total 52 mutants were generated for linear-SVM and 50 for RBF-SVM. Of these mutants, we removed those which throw a run-time exception or those which do not affect the program's output (e.g., changes to the `print' function). This resulted in 6 mutants of interest (the mutants were the same for both linear \& rbf kernel)\footnote{The entire set of mutants and analysis is here: \url{https://github.com/verml/VerifyML/tree/master/ML_Applications/SVM/Mutants}}. All the 6 mutants were of a similar kind and read the label of a data instance (which denotes the class) from a wrong column of the .csv data file. One of the mutants is shown in Figure \ref{svmMutant}.

\begin{figure}[H]
	\begin{verbatim}
   digits_target = digits[:,(-1)].astype(np.int)
											
   digits_target = digits[:,1].astype(np.int)
	\end{verbatim}
\caption{Original Code (top) and Mutant l2 (below). This mutant will read the wrong column of the data for the label.}
\label{svmMutant}
\end{figure}

\subsubsection{Results}
We took the training \& test data available with the application and generated new datasets as per the MRs. Each of the 6 mutants was then executed with the original data and the new data. When the outputs (viz., class label and the distance of a data instance from the decision boundary) from the mutants did not match as per the MR, we report the mutant as killed. The results of running the MRs over the mutants are shown in Table \ref{svmLinRBFResults}.

The results show that all the six mutants of linear-SVM and RBF-SVM were caught through the MRs. In particular, MR-1 (permutation of input features) was sufficient to catch all the mutants. This is because, all the 6 mutants generated by the MutPy tool correspond to  incorrect handling of input data (e.g., the mutant l2 as shown in Figure \ref{svmMutant}). MR-1 changed the input feature which caused the SVM to think the class label has changed and thus gave different outputs. For precisely the same reason, MR-3 (shifting the features by a constant) caught all the mutants as well.

\begin{table}[H]
	\caption{MRs applied on the mutants of linear-SVM \& RBF-SVM. $\checkmark$ denotes the mutant was killed.}
	\begin{tabu}{|X[l]|m{0.3cm}|m{0.3cm}|m{0.3cm}|m{0.3cm}|m{0.3cm}|m{0.3cm}|}
	\hline
	\multirow{2}{*}{MR} & \multicolumn{6}{c|}{Mutant Num of linear-SVM} \\
	\cline{2-7}
	& l2 & l5 & l8 & l11 & l22 & l31 \\
	\hline
	MR-1 (permute features) & \checkmark &\checkmark & \checkmark&\checkmark &\checkmark &\checkmark \\
	\hline
	MR-2 (re-order train instances) & & &  & & & \\
	\hline
	MR-4 (scale test features) & & & & & & \\
	\hline
	\multirow{2}{*}{} & \multicolumn{6}{c|}{Mutant Num of RBF-SVM} \\
	\cline{2-7}
		& r2 & r5 & r8 & r11 & r22 & r31 \\
	\hline
	MR-1 (permute features) & \checkmark &\checkmark &\checkmark &\checkmark &\checkmark &\checkmark \\
	\hline
	MR-2 (re-order train instances) & & & & & & \\
	\hline
	MR-3 (shift train \& test features) & \checkmark &\checkmark &\checkmark &\checkmark &\checkmark &\checkmark \\
	\hline
	\end{tabu}
	\label{svmLinRBFResults}
\end{table}

\subsection{Results from Metamorphic Testing of Application 2}
\begin{figure*}
	\subfloat[MR-1 on Mutant c43]{
	\includegraphics[width=0.23\textwidth]{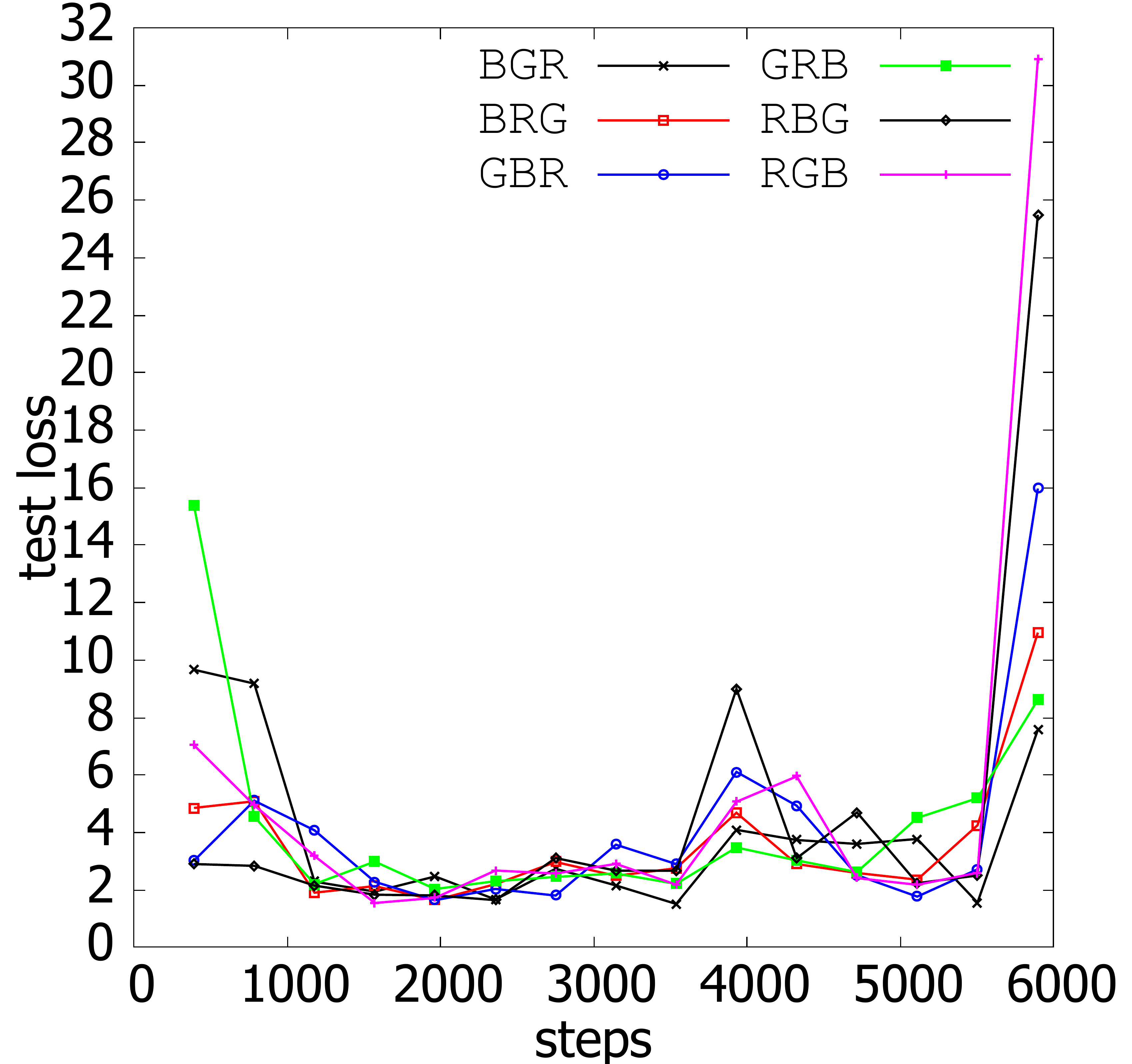}}%
	\hspace{8pt}%
	\subfloat[MR-1 on Mutant c44]{
	\includegraphics[width=0.23\textwidth]{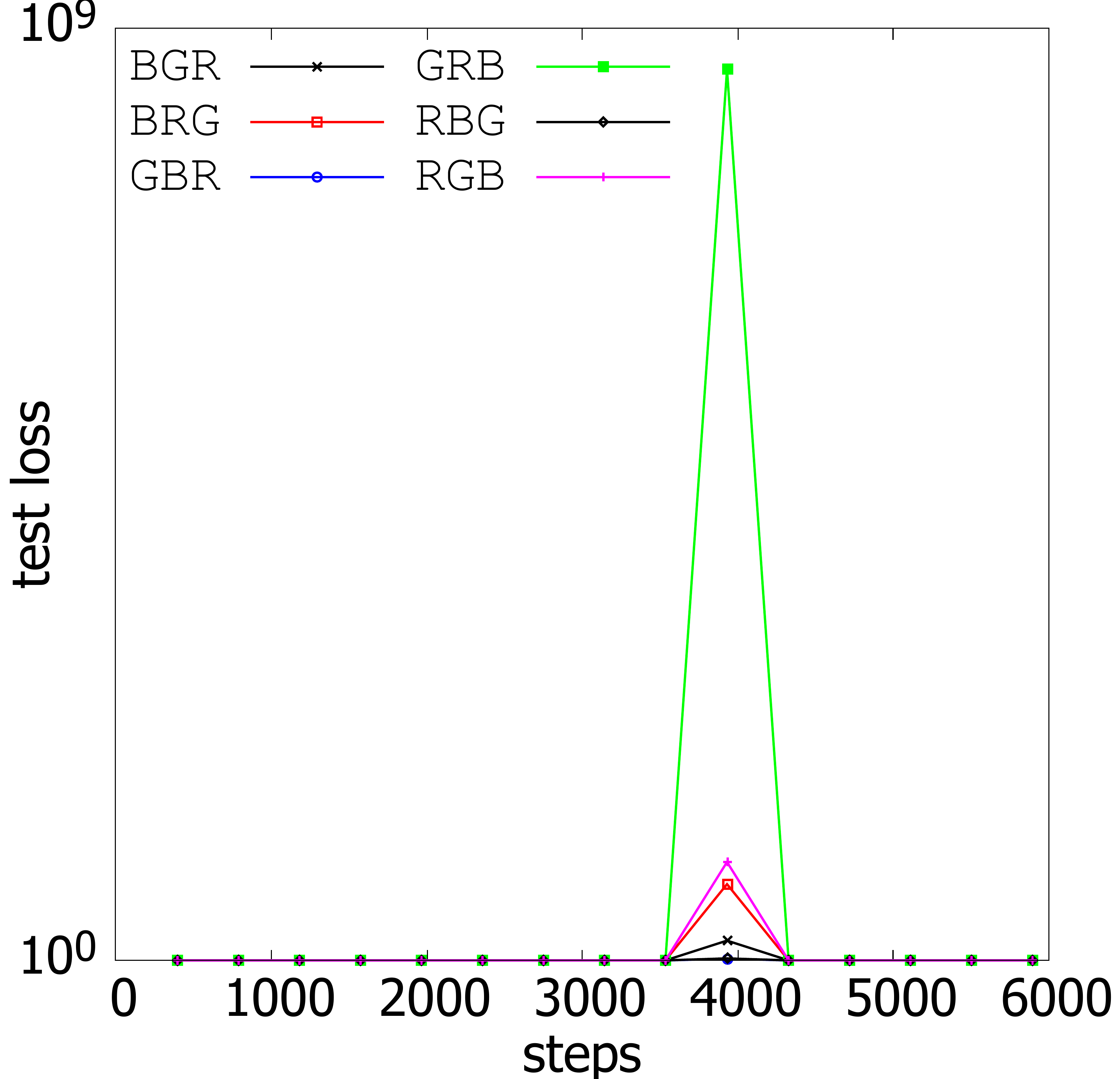}}%
	\hspace{8pt}%
	\subfloat[MR-1 on Mutant c221]{
	\includegraphics[width=0.23\textwidth]{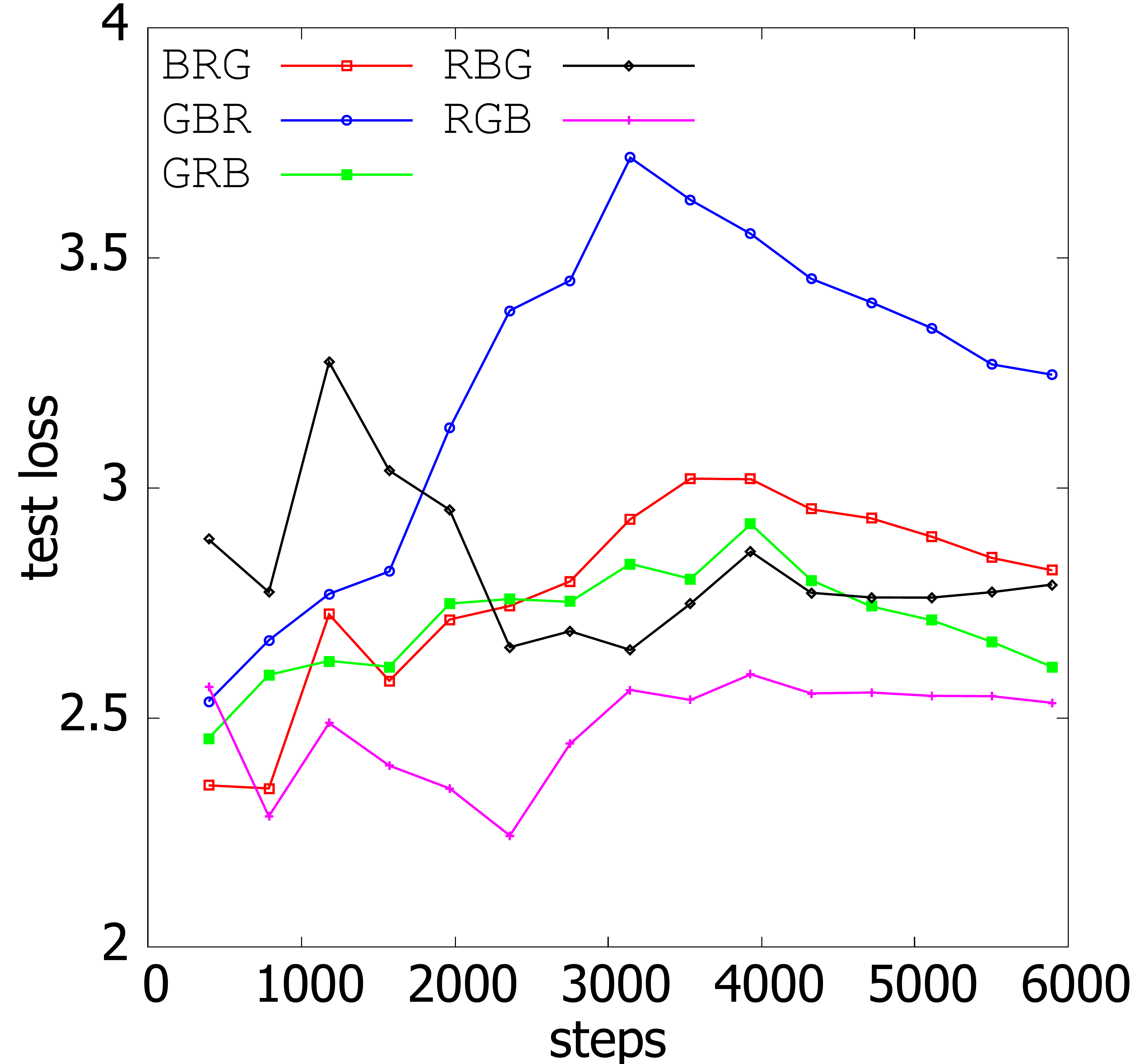}}%
	\hspace{8pt}%
	\subfloat[MR-1 on Mutant c50]{
	\includegraphics[width=0.23\textwidth]{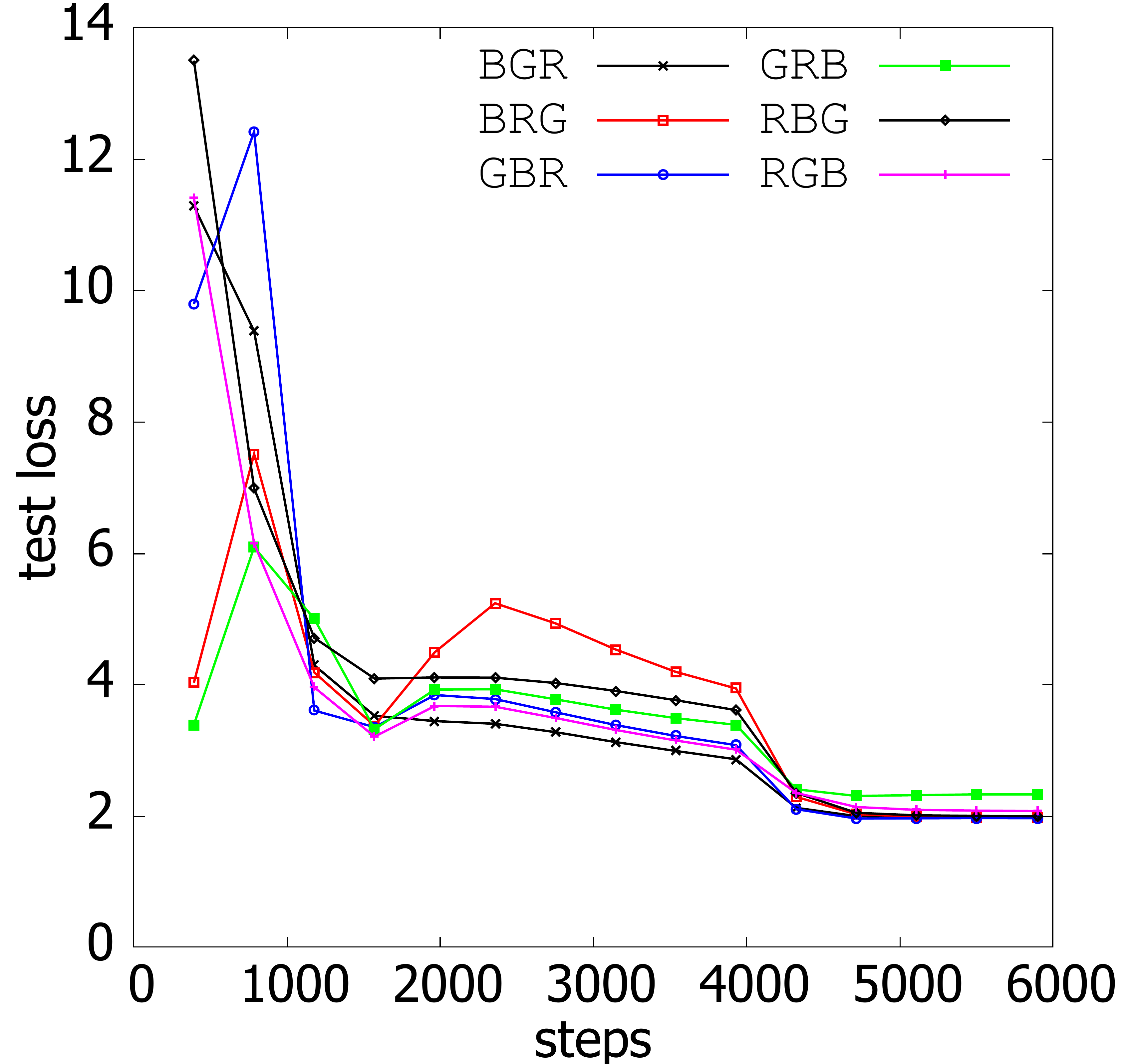}}%
\\%
	\subfloat[MR-2 on Mutant c29]{
	\includegraphics[width=4cm]{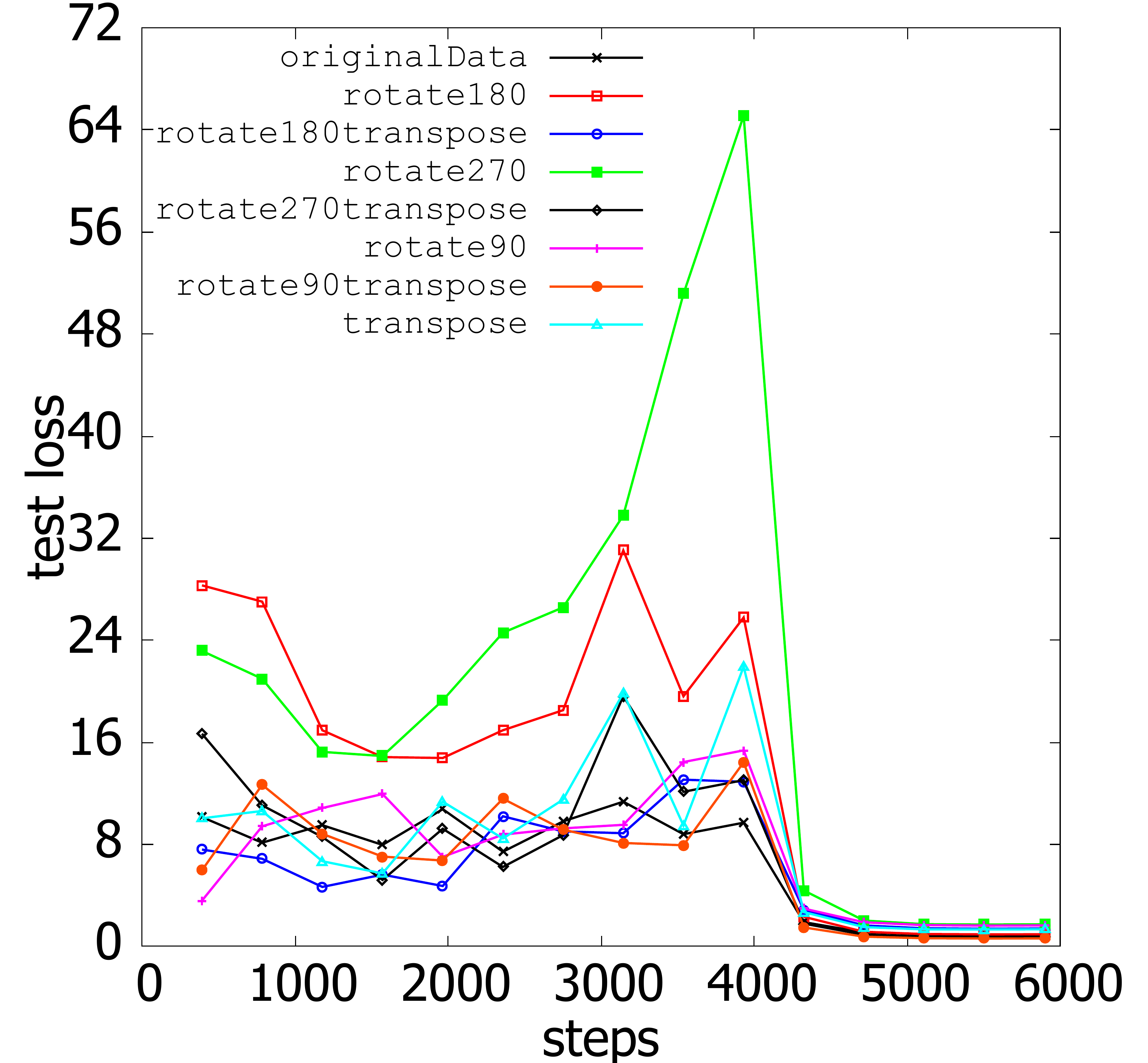}}%
	\hspace{8pt}%
	\subfloat[MR-2 on Mutant c31]{
	\includegraphics[width=4cm]{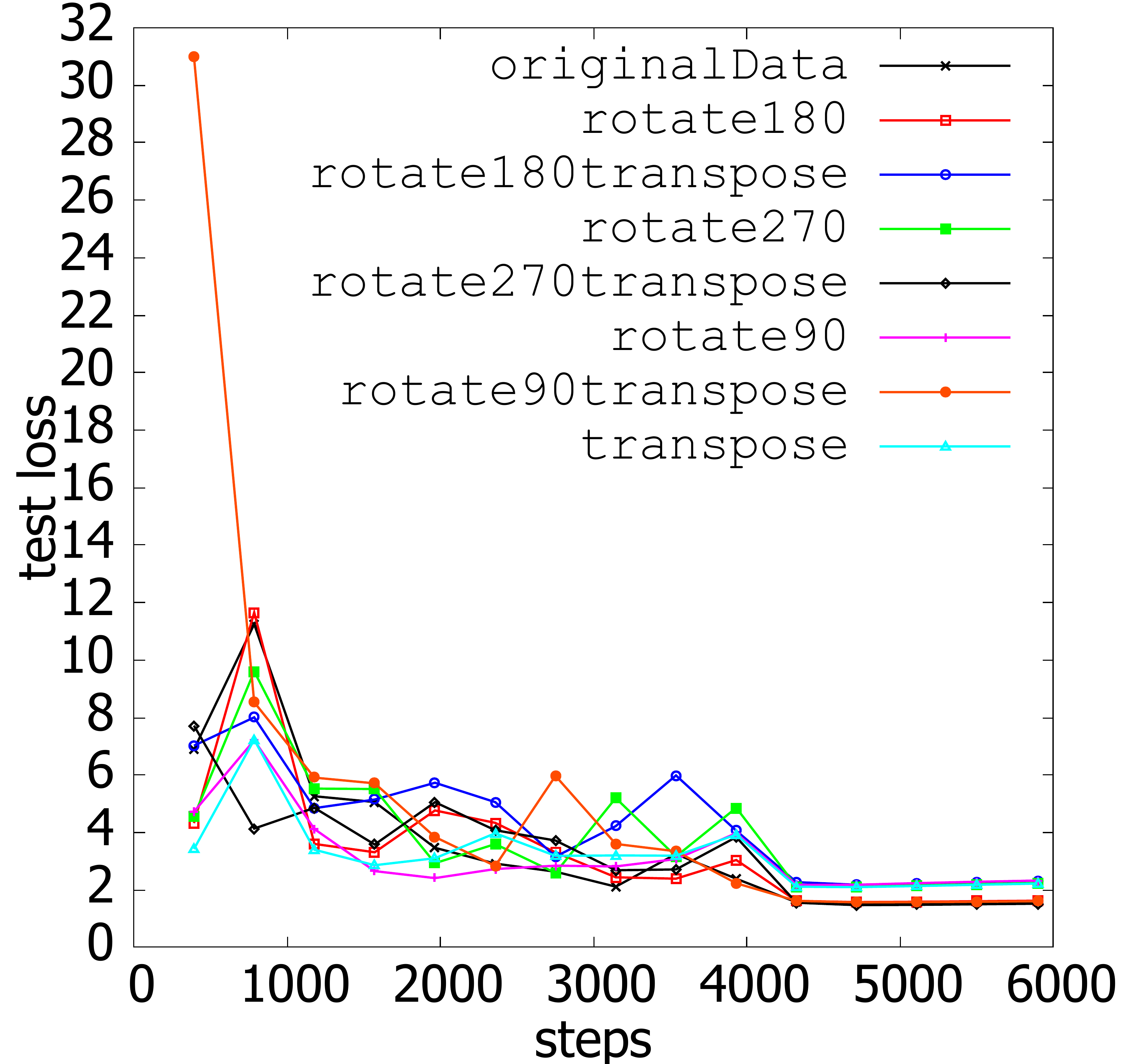}}%
	\hspace{8pt}%
	\subfloat[MR-2 on Mutant c49]{
	\includegraphics[width=4cm]{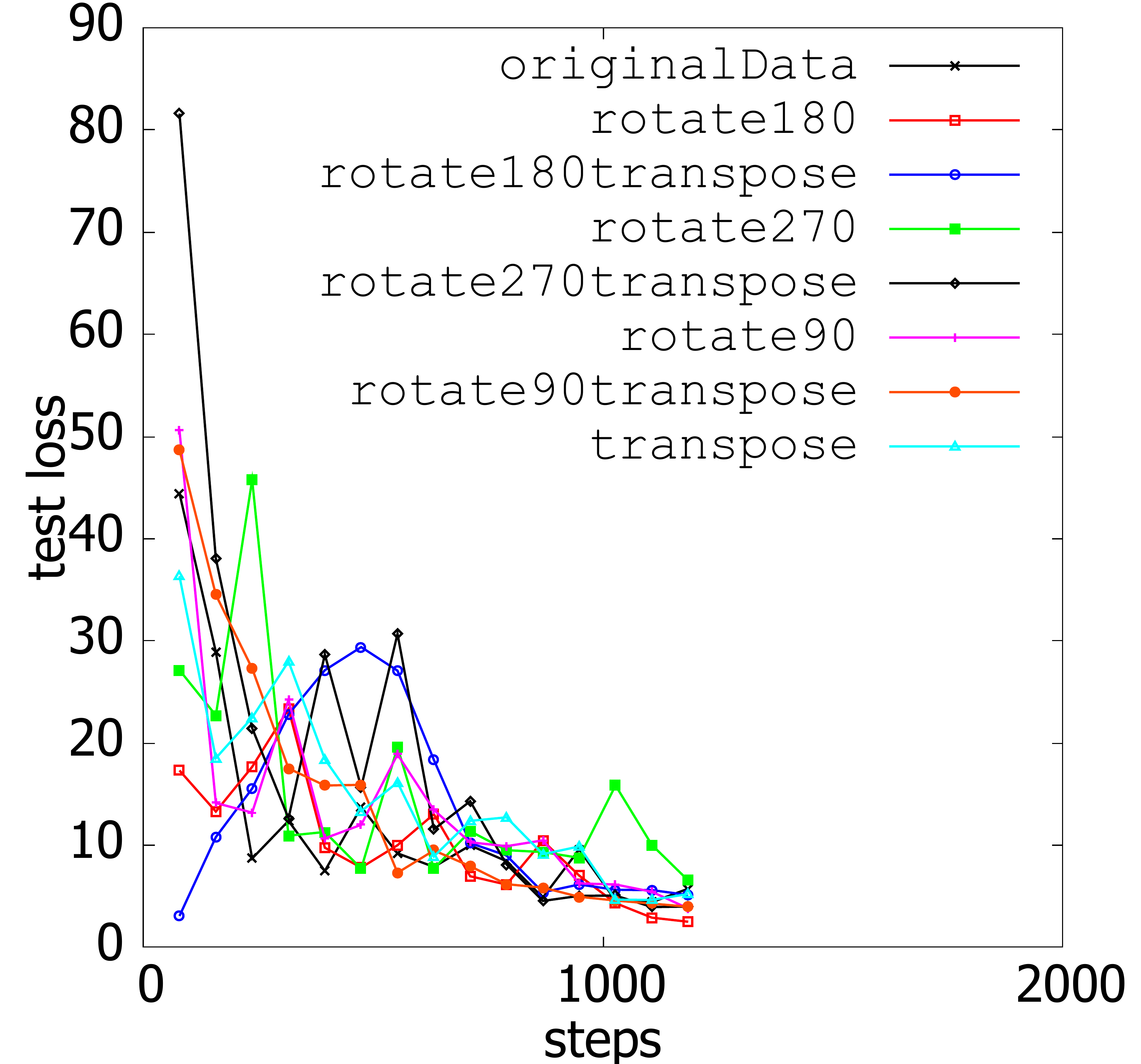}}%
	\hspace{8pt}%
	\subfloat[MR-2 on Mutant r49]{
	\includegraphics[width=4cm]{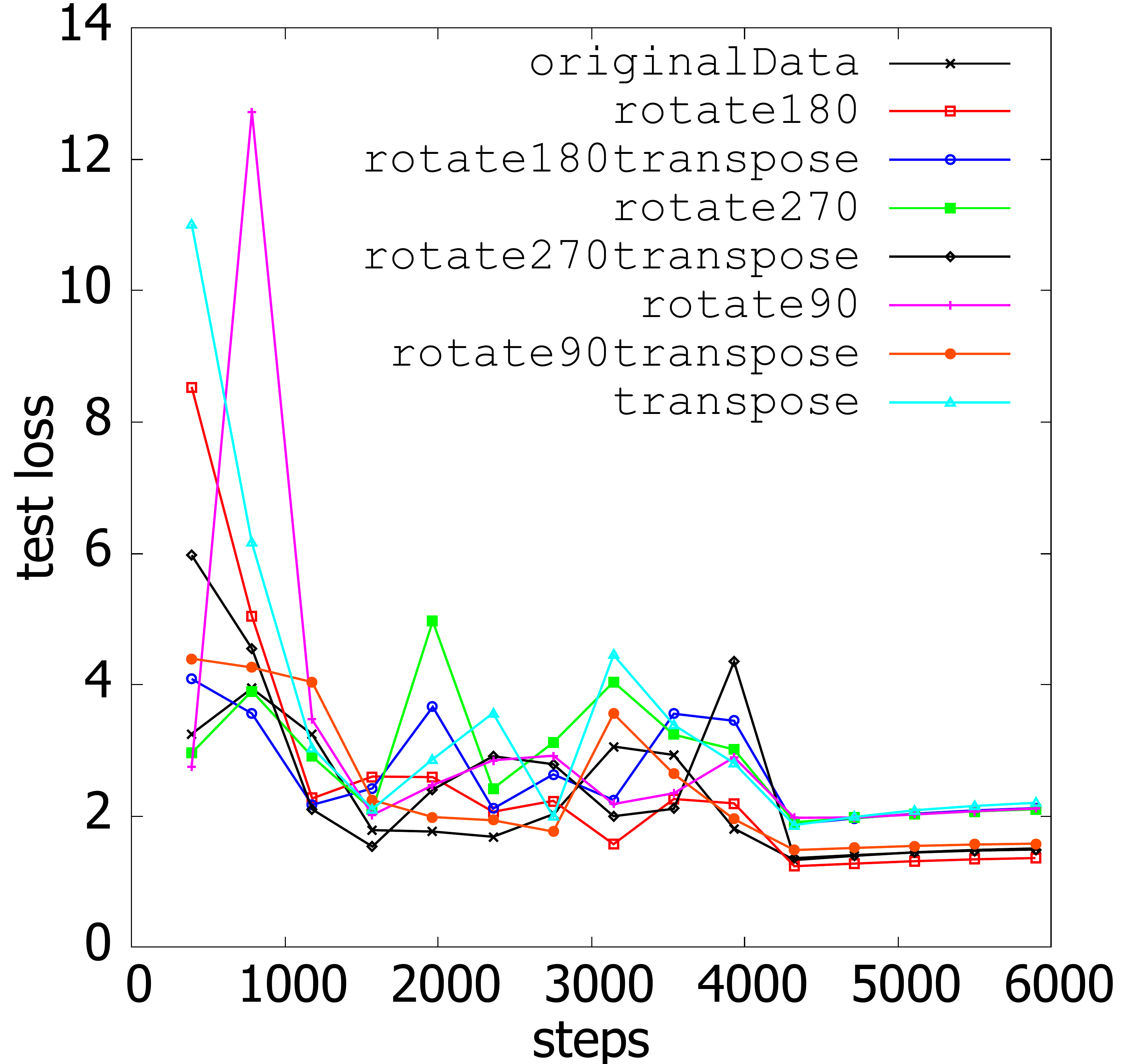}}%
%
	\caption{Test loss when mutants are run with MR-1 (top four graphs) and MR-2 (bottom four graphs). First three plots of each row are mutants that are caught. Clear outliers can be seen. Fourth plot of each row are mutants that were not caught. Please see Appendix B \cite{ourAppendix} for results for all mutants and for a discussion on why some were caught and others weren't.}
	\label{validationLoss_mutants}
\end{figure*}

One of the challenges working with the ResNet application (and possibly deep neural networks) is the stochastic nature of the output where different results are seen for two runs with the same inputs. Such differing outputs is a challenge for Metamorphic Testing, as the MRs compare outputs of subsequent runs. We found stochasticity in the ResNet application to be due to the usage of random seeds. To alleviate this, we updated the code to use a fixed seed for each run. Fixing the seed made the application deterministic when run on a CPU, unfortunately, the application was still stochastic when executed on a GPU. Appendix C \cite{ourAppendix} provides details on the amount of variance seen on the GPU. We could not determine the cause for the non-determinism on the GPU, but it appears to be an issue with the NVidia CUDA libraries \cite{nonDeter1} \cite{nonDeter2}. Thus, all our experiments were run on a CPU. Whether our results can be replicated on a GPU needs to be studied further.

The ResNet application trains by making multiple passes over the training data (called epochs). 150 epochs are made to complete the training. Completing 150 epochs of training on the entire CIFAR-10 data takes around 105 hours on an Intel i5 CPU. Therefore, we carefully selected $10\%$ of the CIFAR-10 data maintaining the data/class distribution. This resulted in 5,029 training and 1010 test instances. On this shortened data, 150 epochs of training takes approximately 10 hours. Our work in this paper has executed 392 experiments of 150 epochs (or 163 days of compute). The experiments were executed on Intel i5 CPU running Windows 10 with Tensorflow 1.7.

\subsubsection{Creating the Mutants}
The ResNet application code is written in two files - \texttt{cifar10.py} and \texttt{resnet.py}. Using MutPy tool \cite{mutpy}, mutants were created for both files. In total, $459$ mutants were created. We analyzed each mutant and discarded those that give a run-time exception, those that do not change the program's output (because they are syntactic changes in code that is never reached) and those that were changes to the hyper-parameters (like the learning rate). This reduction led to 16 valid mutants \footnote{The entire set of mutants and the analysis is available here:\url{https://github.com/verml/VerifyML/tree/master/ML_Applications/CNNs/Mutants}}. The types of valid mutants created is shown in Table \ref{typeOfMutants}. This gives an idea of the typical errors that may be created while writing deep learning applications in Python. Figure \ref{mut29} shows a mutant of type `Change the loss function'.

\begin{table}
\caption{Types of Mutants created for ResNet application.}
	\begin{tabu}{|X[l]|m{3cm}|}
	\hline
	Mutant Type & Num. of Mutants\\
	\hline
	Reduce the training data files & 3  \\
	\hline
	Change the loss function & 4  \\
	\hline
	Changes to the learning rate decay & 3 \\
	\hline
	Interchange training and testing & 2 \\
	\hline
	Change the architecture of ResNet & 2 \\
	\hline
	Pad the wrong channels & 2 \\
	\hline
	Total & 16 \\
	\hline
	\end{tabu}
\label{typeOfMutants}
\end{table}

\begin{figure}[H]
	\begin{verbatim}
       loss = cross_entropy + _WEIGHT_DECAY
											
       loss = cross_entropy - _WEIGHT_DECAY
	\end{verbatim}
\caption{Original Code (top) and Mutant c29 (below). This mutant will explicitly attempt to overfit on the data.}
\label{mut29}
\end{figure}

\subsubsection{Results}
Figure \ref{validationLoss_mutants} shows the plot of the test loss for a few mutants when run against MR-1 \& MR-2. The plots capture the test loss that is output by the application after around every 300 steps of training. Plots of all mutants are available in Appendix B \cite{ourAppendix}. From Figure \ref{validationLoss_mutants}, we can clearly see strong outliers for some of the mutants. However, we see no such outliers on the original (non-buggy) code (Figure \ref{validationLoss_originalCode}) across datasets and architectures. To catch a mutant, we computed the standard deviation ($\sigma$) of the test loss at every step (across all variants of a MR) and considered the maximum deviation $\sigma_{max}$ across all steps. For example, $\sigma$ for MR-1 measures how deviant are the test losses of a mutant for all channel variants, RGB, BGR, etc., at a given step in training. For MR-1 \& MR-2 we term the mutant as caught if this $\sigma_{max}$ is beyond a threshold (i.e. there is strong evidence that the variants of the MRs are not behaving alike). We used a threshold of 9 to report the results in Table \ref{resNetResults}, however, the results showed that the number of mutants killed would be same for any threshold between 5 to 9 (an indication of the large outliers seen in output of mutants). The test only MRs (MR-3 \& MR-4) were also able to catch mutants. This is very useful since the MRs can be run very quickly since they work on an already trained model. 

The MRs are not able to catch any of the r* mutants since all of them pertain to change in the architecture of the ResNet. For example, mutant r48 pads the depth channel of the image (thereby increasing the number of tunable parameters and increasing the capacity of the network). Such mutants on changes in architecture did not show any noticable change in the outputs. It would be extremely interesting to explore whether any MRs can indeed catch such mutants. For a discussion on why mutants are (not-)caught, please refer to Appendix B \cite{ourAppendix}. 

\subsection{Discussion}
Our experimental results show that the defined MRs for the SVM application is able to catch all of the 12 mutants (6 each for linear \& RBF kernels) while the MRs defined for the ResNet application is able to catch 8 out of 16 mutants (for a total of 20/28 or 71\%), which is promising. Metamorphic testing opens an interesting approach to test ML applications where one takes into account the characteristics of the datasets \& the learning algorithm to define relations/properties that are expected to hold. This eases the burden on the testing team from creating large data sets for validation. Instead, the testing team can focus on generating generic metamorphic relations for a class of learning algorithms and datasets (e.g., for image classification, text analytics, etc.). Due to the computational complexity (long training times) of ResNet application, we selected a subset of data to validate our approach. It would make an interesting proposition to replicate the results on the full data set (although we expect to see a similar behavior). Furthermore, we would like to assess the strength of the defined MRs for other deep learning architectures.

\begin{table}[!htb]
	\caption{MRs applied on mutants of ResNet. Values in brackets report $\sigma_{max}$. In total 8 out of 16 (50\%) of bugs were caught.}
	\begin{tabu}{|X[-0.9,c,m]|X[c]|X[-1.3,c]|X[c]|X[-0.8,c]|}
	\hline
	Mutant & MR-1 (permute RGB) & MR-2 (permute CONV order) & MR-3 (normalize data) & MR-4 (scale data) \\
	\hline
	c9 &  & & & \\
	\hline
	c29 &  &\checkmark (18.1) & & \\
	\hline
	c30 &  & & & \\
	\hline
	c31 &  &\checkmark (9.1) & & \\
	\hline
	c32 &  &\checkmark (9.1) & & \\
	\hline
	c43 & \checkmark (9.6) & \checkmark (11.5) & & \\
	\hline
	c44 & \checkmark (27.4) & \checkmark (9.2) & & \\
	\hline
	c45 &  & & & \\
	\hline
	c49 & \checkmark (23.3) &\checkmark (23.8) & &\checkmark \\
	\hline
	c50 &  & &\checkmark &\checkmark \\
	\hline
	c116 &  & & & \\
	\hline
	c221 & \checkmark ($\infty$) & & & \\
	\hline
	r6 &  & & & \\
	\hline
	r48 &  & & & \\
	\hline
	r49 &  & & & \\
	\hline
	r67 &  & & & \\
	\hline
	\end{tabu}
	\label{resNetResults}
\end{table}

\section{Conclusion} \label{conclusion}
In this paper we have investigated the problem of identifying implementation bugs in machine learning based applications. Current approaches to test an ML application is limited to `validation', where the tester acts as a human-oracle. This is because ML applications are complex and verifying its output w.r.t the specification is extremely challenging. Our solution approach is based on the concept of Metamorphic Testing where we build multiple relations between subsequent outputs of a program to effectively reason about the correctness of the implementation. We have developed such relations for two image classification applications - one using a Support Vector Machine (SVM) classifier and another application using Deep Learning based Convolutional Neural Network. Experimental results showed that, on average, 71\% of the implementation bugs were caught by our method and gives impetus for further exploration.

\bibliographystyle{ACM-Reference-Format}
\balance
\bibliography{ref} 

\end{document}